\documentclass[%
preprint,
nofootinbib,
amsmath,amssymb,
aps,
floatfix, 
]{revtex4-2}

\usepackage{lineno}[displaymath, mathlines]
\usepackage{graphicx}
\usepackage{dcolumn}
\usepackage{bm}
\usepackage{float,tabularx, multirow, appendix, booktabs}
\usepackage{hyperref}
\usepackage{multirow}
\usepackage{comment}

\usepackage[noabbrev,capitalise]{cleveref}
\crefname{equation}{equation}{equations}
\Crefname{equation}{Equation}{Equations}
\usepackage{xcolor}
\usepackage{appendix}
\usepackage[linesnumbered,ruled]{algorithm2e}

\usepackage{soul}
\setstcolor{blue}

\usepackage{tikz}
\usetikzlibrary{shapes,arrows}

\soulregister{\fh}{1}

\newcounter{savesection}
\newcounter{apdxsection}
\newcommand\unappendix{\par
  \setcounter{apdxsection}{\value{section}}%
  \setcounter{section}{\value{savesection}}%
  \setcounter{subsection}{0}%
  \gdef\thesection{\@arabic\c@section}}

\begin{document}

\preprint{}

\title{Data-driven prediction and control of extreme events \\in a chaotic flow}

\author{Alberto Racca}
\affiliation{%
University of Cambridge, Department of Engineering, Trumpington Street, CB2 1PZ, Cambridge, United Kingdom
}
\author{Luca Magri}%
\email{lm547@cam.ac.uk}
\affiliation{%
Imperial College London, Aeronautics Department, South Kensington Campus
London SW7 2AZ, United Kingdom
}%

\affiliation{%
University of Cambridge, Department of Engineering, Trumpington Street, CB2 1PZ, Cambridge, United Kingdom
}%

\affiliation{%
The Alan Turing Institute,  96 Euston Road,  NW1 2DB, London, United Kingdom
}%

\affiliation{%
Institute of Advanced Study, TU Munich,  Lichtenbergstrasse 2 a, 85748 Garching bei M\"unchen, Germany (visiting)
}

\date{\today}


\begin{abstract}


An extreme event is a sudden and violent change in the state of a nonlinear system.
In fluid dynamics, extreme events can have adverse effects on the system's optimal design and operability, which calls for accurate methods for their prediction and control. 
In this paper, we propose a data-driven methodology for the prediction and control of extreme events in a chaotic shear flow. 
The approach is based on echo state networks, which are a type of reservoir computing that learn temporal correlations within a time-dependent dataset. 
%
%
%
The objective is five-fold. 
First, we exploit {\it ad-hoc} metrics from binary classification to analyse 
(i) how many of the extreme events predicted by the network actually occur in the test set (precision), and 
(ii) how many extreme events are missed by the network (recall). We apply a principled strategy for optimal hyperparameter selection, which is key to the networks' performance. 
Second, we focus on the time-accurate prediction of extreme events. 
We show that
 echo state networks are able to predict extreme events well beyond the predictability time, i.e., up to more than five Lyapunov times. 
Third, we focus on the long-term prediction of extreme events from a statistical point of view. 
By training the networks with datasets that contain non-converged statistics, we show that the networks are able to learn and extrapolate the flow's long-term statistics. In other words, the networks are able to extrapolate in time from relatively short time series. 
Fourth, we design a simple and effective control strategy to prevent extreme events from occurring.
The control strategy decreases the occurrence of extreme events up to one order of magnitude with respect to the uncontrolled system.
Finally, we analyse the robustness of the results for a range of Reynolds numbers. We show that the networks perform well across a wide range of regimes. 
This work opens up new possibilities for the data-driven prediction and control of extreme events in chaotic systems.

\end{abstract}

\maketitle
  
\clearpage
\section{Introduction}


Extreme events are sudden and violent changes of an observable.
These events arise in a variety of fluid dynamic systems in the form of oceanic rogue waves \cite{farazmand2017variational, majda2019statistical}, dissipation events in turbulence \cite{yeung2015extreme, blonigan2019extreme}, thermoacoustic bursts in aeronautical engines  \cite{lieuwen2005combustion} and atmospheric blocking events in weather forecasting  \cite{tibaldi1990operational}, to name a few. Because of their large amplitudes, extreme events can become problematic in many engineering applications because they may lead to component failure \cite{hassanaly2021classification}. Because of the potentially negative consequences of extreme events, their prediction and suppression is an active area of research \cite{ghil2011extreme,Rahmstorf2011,farazmand2019extreme,sapsis2021statistics}.\\

There are two ways to study extreme events: 
(i) computing their statistics and (ii) time-accurately predicting the incoming events. With a statistical approach, the goal is to predict the probability of the occurrence of an extreme event. To do so, methods such as extreme value theory \cite{lucarini2016extremes}, large deviation theory \cite{varadhan2010large} and tailored Monte Carlo algorithms \cite{Margazoglou2019} are used to compute the heavy tail of the probability density functions that describe the events. With a time-accurate approach, the goal is to identify precursors, which are quantities that indicate at the current time the occurrence of an  extreme event in the near future \cite{cavalcante2013predictability,farazmand2017variational}. Although the time-accurate prediction is mathematically possible in deterministic  systems, it is hindered by the ‘‘butterfly effect’’: in chaotic dynamics the system is predictable only up to a time scale, known as the predictability time \cite{lorenz1963deterministic}. One simple scale for the predictability time is the inverse of the dominant Lyapunov exponent, also known as the Lyapunov time, which is typically a small characteristic scale of the flow \cite{boffetta2002predictability}.\\

In turbulence and chaotic flows, extreme events have been characterized  with precursors \cite{farazmand2019extreme}. The authors of \cite{babaee2016minimization} developed an optimal time-dependent reduced-order model, which consists of a basis that at any point of the trajectory spans the most unstable directions characterized by the largest finite-time growth of perturbations. The eigenvalues of the linear operator describing the evolution of perturbations in the reduced-order model performed well in predicting dissipation events in forced periodic turbulence \cite{farazmand2016dynamical}. However, computing the elements of the basis may be computationally prohibitive for systems with a large number of degrees of freedom \cite{sapsis2018new}. The authors of \cite{farazmand2017variational} developed a variational method to find the optimal condition along the attractor that maximizes the instantaneous growth of the observable exhibiting extreme events, which was applied to predict rogue waves and forced periodic turbulence. These methods are \emph{model-based}, because they rely on the perfect knowledge of the governing equations. \\

A different approach to the prediction of extreme events is \emph{data-driven}. 
Given some time series (data), we wish to predict the future evolution of the system, usually without any knowledge of the governing equations. In the realm of data-driven techniques for predicting time series, Recurrent Neural Networks (RNNs) \cite{rumelhart1986learning}, and their variants, are the state-of-the-art machine learning architectures. They are designed to have memory of the past history of the system through an internal state, which enables the network to learn temporal correlations within the data. Because of their design, RNNs typically outperform other common machine learning architectures, such as feed-forward neural networks, in time series prediction \cite{chattopadhyay2019data}. In fluid dynamics, RNNs have been successfully applied for different goals such as closure for reduced-order modelling \cite{wan2018data}, prediction of flows past bluff bodies \cite{hasegawa2020machine}, replication of chaotic flow statistics \cite{snrinivasan2020} and nonlinear control for gliding \cite{novati2019}, to name a few. \\

Among recurrent neural networks, Echo State Networks (ESNs) \cite{jaeger2004harnessing,maass2002real}, which are a class of reservoir computing, are a versatile architecture for the prediction of chaotic dynamics. They have been proven to be universal approximators \cite{GRIGORYEVA2018}, which can infer the invariant measure of  dynamical systems \cite{huhn2020learning} under non-stringent assumptions of ergodicity~\cite{hart2021echo}. ESNs perform as well as other architectures such as Long Short-Term Memory (LSTM) Networks \cite{hochreiter1997long}, whilst requiring less computational resources for training \cite{chattopadhyay2019data,vlachas2020backpropagation}. Although training echo state networks is a straightforward task as compared to other networks, such as LSTMs, their performance is highly sensitive to the selection of hyperparameters \cite{lukovsevivcius2012practical,racca2021robust}. To improve the performance of the networks, prior knowledge of the governing equations has been embedded in the hybrid \cite{pathak2018hybrid} and physics-informed \cite{DOAN2020} architectures. In fluid dynamics, echo state networks have been employed to (i) learn \cite{huhn2020learning} and (ii) optimize ergodic averages in thermoacoustic oscillations \cite{huhn2021gradient}; (iii) time-accurately \cite{doan2021short} and (iv) statistically predict extreme events in chaotic flows \cite{doan2021short, racca2022statistical}. In particular, \citet{doan2021short} studied the effect of including prior knowledge of the governing equations in a chaotic shear flow model to improve the prediction of extreme events when only a small training set is available. Although leveraging physics knowledge improves performance, information about the governing equations is not always available,  and the networks are sensitive to the selection of hyperparameters \cite{racca2021robust}.  \\ 

Once an upcoming extreme event is accurately  predicted, the next objective is to suppress the event, i.e., to prevent the event from occurring. This is usually achieved via a control strategy that is based on the prediction of the event. The authors of \cite{cavalcante2013predictability}  used information from an empirical precursor to suppress extreme events in a specifically designed six-dimensional system of electronic circuits; whereas \cite{pyragas2020using} apply ESNs to suppress local extreme events in a system of coupled oscillators; and \cite{farazmand2019closed} computed the most unstable initial conditions by solving an optimization problem to successfully suppress extreme events through closed-loop control in two-dimensional turbulence. 
\\

The objective of the paper is five-fold. 
First, we define a set of metrics to evaluate the performance of the networks for the time-accurate and statistical prediction of extreme events. The importance of a reliable hyperparameter selection is emphasized. 
Second, we investigate the time-accurate prediction of extreme events using echo state networks for times larger than the Lyapunov time. 
Third, we analyse the networks' capability to extrapolate statistical knowledge of the system with respect to the available data used for training. 
Fourth, we exploit the time-accurate prediction of extreme events to actively suppress the events, with a simple and effective control strategy. 
Fifth, we investigate the robustness of the networks in a wide range of Reynolds numbers.

The paper is organised as follows. 
Section \ref{sec:MFE} discusses the qualitative model of chaotic shear flow and the generation of the dataset. 
Section \ref{sec:ESN} describes the echo state network architecture. 
Section \ref{sec:time-acc} defines the metrics and analyses the time-accurate prediction of extreme events. 
Section \ref{sec:stats} analyses the statistical prediction of extreme events with respect to the available datasets. 
Section \ref{sec:control} proposes an open-loop control strategy based on the time-accurate prediction of extreme events.  
Finally, section \ref{sec:concl}
summarizes the results of this study and discusses current/future work.\\

\section{A qualitative model of chaotic shear flow}

\label{sec:MFE}
\begin{figure}[H]
\centering
\includegraphics[width=1\textwidth]{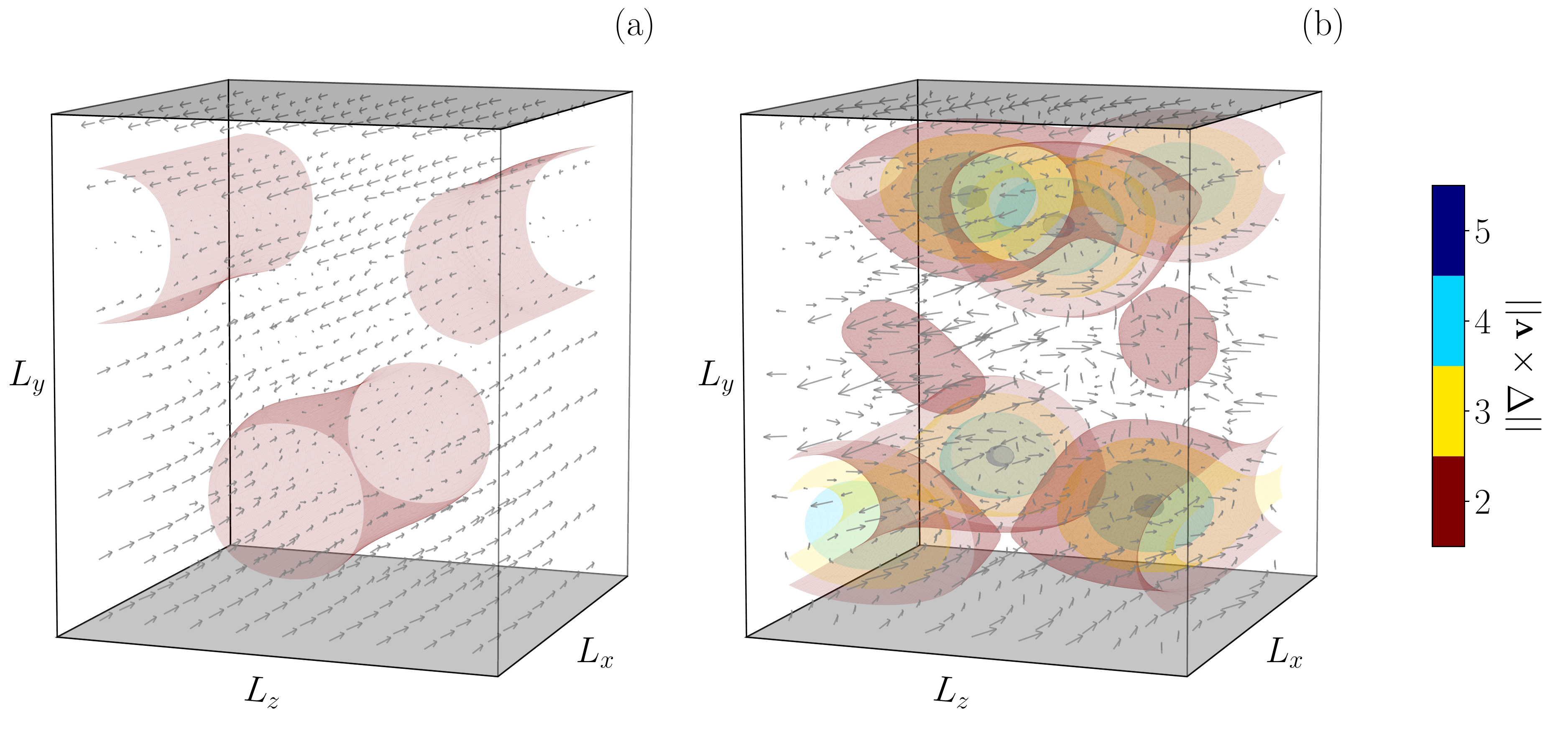}
\caption{Vorticity isosurfaces and velocity flowfield (a) before and (b) after an extreme event. The laminar structure (a) breaks down into vortices (b).}
\label{fig:3d_wort}
\end{figure}

\begin{figure}[H]
\includegraphics[width=1.0\textwidth]{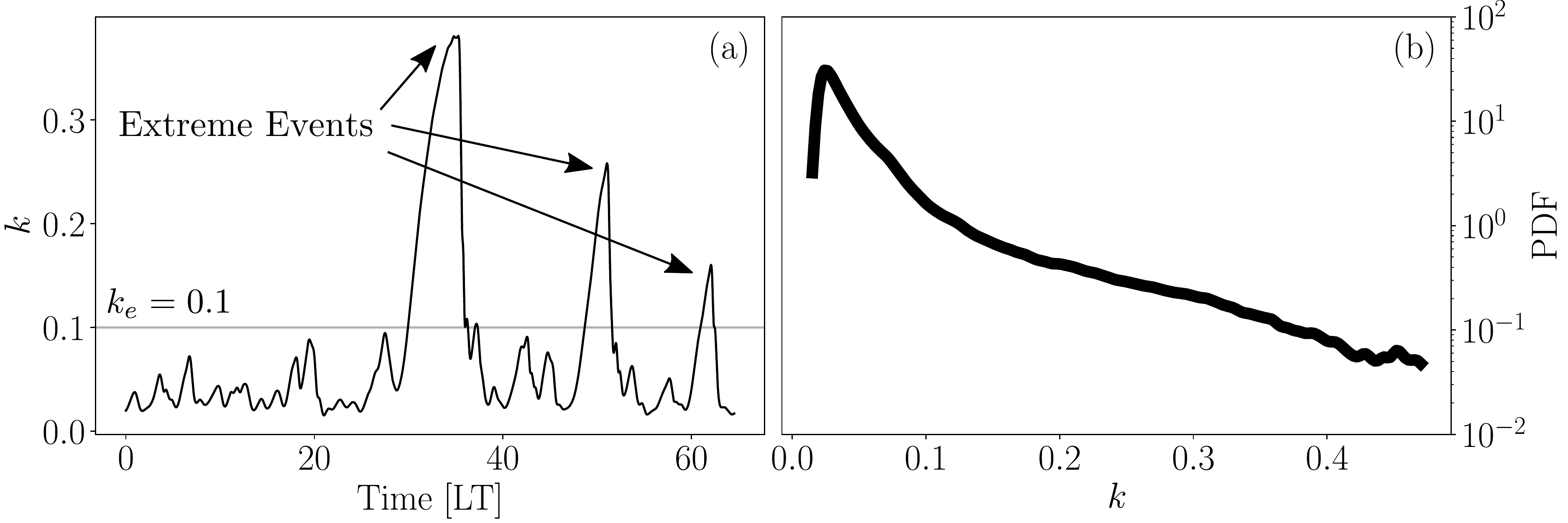}
\caption{(a) Typical evolution of the kinetic energy. Time is expressed in Lyapunov times (LT) (Section~\ref{sec:lyap}). (b) Probability density function of the kinetic energy.}
\label{fig:k_time_pdf}
\end{figure}

\begin{figure}[H]
\centering
\includegraphics[width=1.\textwidth]{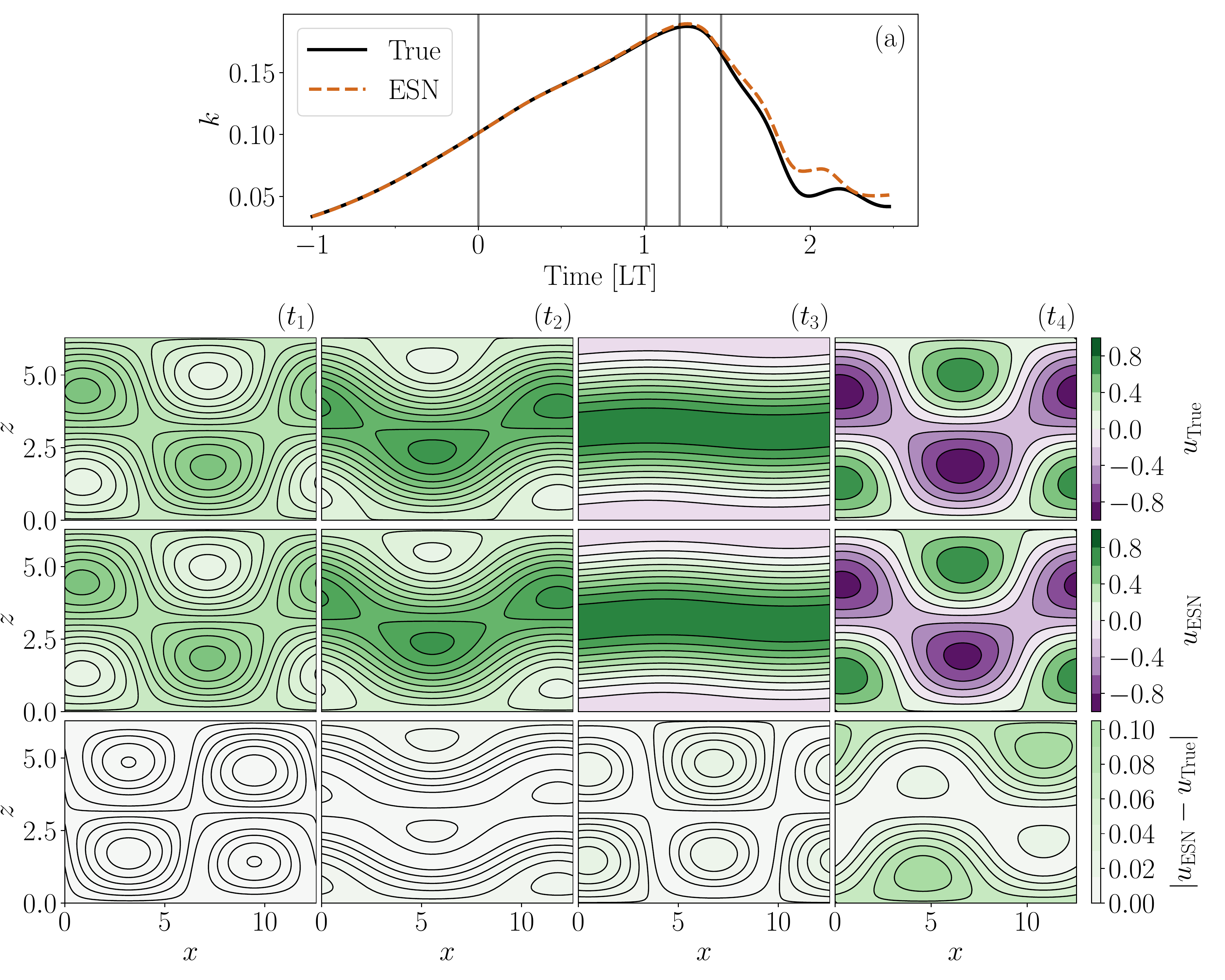}
\caption{Velocity component along $x$ in the midplane during an event in the test set for the data, $u_{\mathrm{True}}$, and an ESN, $u_{\mathrm{ESN}}$. The times, $(t_i)$, of the flowfields are indicated by vertical lines in (a). Times $t_3$ and $t_4$ coincide with $(a)$ and $(b)$ of \cref{fig:3d_wort}. The maximum of the absolute error, $|u_{\mathrm{ESN}}-u_{\mathrm{True}}|$, is approximately $5\%$ the range of the flowfield.}
\label{fig:flows}
\end{figure}

We consider a qualitative model of a shear flow between infinite plates subjected to a sinusoidal body forcing~\cite{moehlis2004low, waleffe1997self}. 
For brevity, this model is referred to as ``MFE''. 
Because the MFE  captures qualitatively nonlinear phenomena, such as relaminarization and turbulent bursts, it has been employed to qualitatively study turbulence transition \cite{skufca2006edge, Eckhardt2007} and predictability of chaotic flows \cite{doan2021short, snrinivasan2020}. 
The dynamics are governed by the non-dimensional Navier-Stokes equations for forced incompressible flows
\begin{equation}
    \frac{d\mathbf{v}}{dt} = - (\mathbf{v}\cdot\nabla)\mathbf{v} - \nabla p + \frac{1}{\mathrm{Re}}\Delta \mathbf{v}  + \mathbf{F}(y),
    \label{eq:NS}
\end{equation}
where $\mathbf{v}=(u,v,w)$ is the velocity, $p$ is the pressure, Re is the Reynolds number, $\nabla$ is the gradient, $\Delta$ is the Laplacian, $\mathbf{F}(y)=\sqrt{2}\pi^2/(4\mathrm{Re})\sin(\pi y/2)\mathbf{e}_x$ is the body forcing along $x$, $y$ is the direction of the shear between the plates,  and $z$ is the spanwise direction (Figure~\ref{fig:3d_wort}). 
We solve the flow in the domain $L_x\times L_y\times L_z$, where the boundary conditions are free slip at $y\pm L_y/2$, and periodic at $x=[0;L_x]$ and $z=[0;L_z]$ (\cref{fig:3d_wort}). The body forcing, $\mathbf{F}(y)$, drives the flow for it to sustain turbulence with homogeneous boundary conditions \cite{waleffe1997self}.
To reduce the partial differential equations~\eqref{eq:NS} to a set of ordinary differential equations, the velocity is projected onto compositions of Fourier modes \cite{moehlis2004low}, $\mathbf{\hat{v}}_i(\mathbf{x})$ as 
\begin{equation}
\label{eq: reduced}
    \mathbf{v}(\mathbf{x},t) = \sum_{i=1}^9a_i(t)\mathbf{\hat{v}}_i(\mathbf{x}).
\end{equation}
The ansatz~\eqref{eq: reduced} is substituted in the Navier-Stokes equations~\eqref{eq:NS}, which are are projected onto the same modes, $\mathbf{\hat{v}}_i(\mathbf{x})$. 
This spawns nine nonlinear ordinary differential equations for the amplitudes, $a_i(t)$, which, in turn, become the unknowns of the system (see supplementary material).
The system displays a chaotic transient, which ultimately converges to the laminar solution $a_1=1,a_2=\dots=a_9=0$. During the transient, the system shows intermittent large bursts in the kinetic energy, which, in the MFE model, can be computed as  
\begin{equation}
    k(t)=\frac{1}{2}\sum_{i=1}^{9}a_i^2(t).
\end{equation}
The bursts in the kinetic energy are the extreme events we wish to predict (Figures~\ref{fig:3d_wort}, \ref{fig:k_time_pdf}(a)). The extreme events are characterized by a heavy tail of the distribution \cite{sapsis2021statistics} (Figure~\ref{fig:k_time_pdf}(b)). 
In this work, we define an event as extreme extreme when the kinetic energy, $k(t)$, exceeds a threshold, $k_e$
\begin{equation}
k(t) \geq k_e,
\end{equation}
where $k_e=0.1$ is the user-defined extreme event's threshold \cite{doan2021short}.
Physically, an extreme event occurs because the stable laminar solution intermittently attracts the chaotic dynamics up to full laminarization. 
In reference to \cref{fig:flows}, during an extreme event, the flow slowly laminarizes ($t_1$ to $t_3$), but the laminar structure suddenly breaks down into vortices ($t_4$). 

\subsection{Lyapunov Exponent}

\label{sec:lyap}

In chaotic systems, almost all nearby trajectories\footnote{Specifically, ``almost'' means that the set of nearby trajectories that will not diverge in time has a Lebesgue-measure of zero. \label{footnote:almost}} diverge at an average exponential rate given by the leading Lyapunov exponent of the system, $\Lambda$ \cite{boffetta2002predictability}. This means that any error in the prediction of the system grows exponentially in time at a rate given by $\Lambda$. Therefore, the leading Lyapunov exponent provides a timescale for assessing the time-accurate prediction of chaotic systems. We use $\Lambda$ to normalize the time units in Lyapunov times, which is defined as $1\mathrm{LT}=\Lambda^{-1}$.
For the calculation of the Lyapunov exponent, we follow the robust procedure used by the authors of~\cite{huhn2019stability} in linear flow analysis. 
We consider a nonlinear autonomous dynamical system in the form of 
\begin{equation} \label{eq:r29ui}
\mathbf{\dot{q}} = \mathbf{f(q)}, 
\end{equation}
where $\mathbf{q}$ is the system's state and $\mathbf{f}$ is a differential operator. 
The state is decomposed as 
\begin{equation} \label{eq:r29ui2}
\mathbf{q}(t)=\mathbf{\overline{q}}(t)+\mathbf{q'}(t), 
\end{equation} 
where $\mathbf{q}'$ is a first-order perturbation that develops on the unperturbed solution, $\mathbf{\overline{q}}(t)$.  
Substituting~\eqref{eq:r29ui2} in~\eqref{eq:r29ui} and collecting the first-order terms yield 
\begin{equation}
    \mathbf{q'}(t)=\mathbf{M}(t,t_0)\mathbf{q'}(t_0), \qquad \qquad \dot{\mathbf{M}}(t,t_0) = \mathbf{J}(t)\mathbf{M}(t,t_0), 
\end{equation}
where $\mathbf{M}(t,t_0)$ is the tangent propagator from $t_0$ to $t$,  $\mathbf{J}(t)\equiv d\mathbf{f(q)}/d\mathbf{q}|_{\mathbf{\overline{q}}(t)}$ is the Jacobian of the system, and $\mathbf{M}(t_0,t_0)=\mathbf{I}$ \cite{huhn2019stability}.
In chaotic systems, perturbations grow in time at an average exponential rate given by the (positive) leading Lyapunov exponent, $\Lambda$, of the system \cite{boffetta2002predictability}
\begin{equation}
    \Lambda = \lim_{t\to\infty}\frac{1}{t}\ln\left(\frac{||\mathbf{q'}(t)||}{||\mathbf{q'}(t_0)||}\right). 
\end{equation}
This means that tangent dynamics are unstable and that the norm, $||\mathbf{q'}(t)|| = ||\mathbf{M}(t,t_0)\mathbf{q'}(t_0)||$, becomes unbounded and causes numerical overflow. To overcome this problem, we use the QR algorithm proposed by \citet{ginelli2007characterizing} to compute the leading Lyapunov exponent. The QR algorithm overcomes the instability of tangent space by performing periodic orthonormalizations of the tangent propagator, $\mathbf{M}(t,t_0)$, thereby avoiding numerical overflow. 
%
Through the QR algorithm, we find the Lyapunov exponent to be $\Lambda=0.0163$\footnote{ The value $\Lambda=0.0163$ is different from the Lyapunov exponent, $\Lambda=0.0296$, computed by \citet{snrinivasan2020}. This difference is because the latter is not the leading Lyapunov exponent, but a local Lyapunov exponent. The local Lyapunov exponent describes the average evolution of perturbations only from one specific initial condition along the trajectory, before nonlinearities saturate the norm of the perturbation. On the other hand, the leading Lyapunov exponent describes the average evolution of perturbations along the entirety of the trajectory. We further verify our results by computing the leading Lyapunov exponent as the average of the exponential growth of perturbations up to nonlinear saturation from different initial conditions along the trajectory, and find, consistently with the QR method, $\Lambda=0.0163$. }, which implies that the Lyapunov time of the system is $1\mathrm{LT}=0.0163^{-1}$.
\newpage 
\subsection{Generation of the dataset}

To generate extreme events, 
we set $L_x=4\pi, L_y=2, L_z=2\pi$ and $\mathrm{Re}=400$ as in~\citet{snrinivasan2020}. 
We generate an ensemble of 2000 time series by integrating the governing equations using a $4$-th order Runge-Kutta scheme with timestep $dt=0.25$. 
A time series of the ensemble is obtained by integrating for 4200 time units the governing equations starting from $\mathbf{a}(0)+0.01\epsilon\mathbf{\hat{e}}_4$, where $\epsilon$ is a random variable sampled from the uniform distribution in $[0,1]$, ${\hat{e}}_4$ is the unit vector pointing in the 4-th direction ($a_4$), and $\mathbf{a}(0) = [1, 0, 0.07066, -0.07076, 0, 0, 0, 0, 0]$. 
We remove the first 200 time units ($\simeq 3 \mathrm{LT}$) of each time series to discard the portion of the time series that is similar among different time series (after some Lyapunov times trajectories diverge one from the other thanks to the butterfly effect). 
Thus, each time series consists of $4000$ time units ($\simeq 65 \mathrm{LT}$). 
\cref{fig:k_time_pdf}(a) shows a typical time series. 

\begin{table}[H]
\centering
\caption{Summary of parameters used for data generation.}
\begin{tabular}{l l l}
\toprule
Parameter $\qquad$ & Explanation & Value \\
\midrule
$L_x$ & Length of the domain in the $x$-direction $\qquad$ & $4\pi$ \\
$L_y$ & Length of the domain in the $y$-direction & $2$ \\
$L_z$ & Length of the domain in the $z$-direction & $2\pi$  \\
Re & Reynolds number & $400$  \\
$dt$ & Integration timestep & $0.25$  \\
$k_l$ & Laminarization threshold & $0.48$  \\
$k_e$ & Extreme event's threshold & $0.1$  \\
 & Number of time series & $1436$  \\
 & Length of each time series & $4000$  \\
 & Number of time series in the training set $\qquad$ & $10$  \\
\toprule
\end{tabular}
\label{tab:dataset generation}
\end{table}
We generate an ensemble of time series because we focus on the chaotic transient dynamics only. 
The ensemble of time series allows us to generate large amount of data, whereas a single long simulation may quickly laminarize, which would provide limited information about the chaotic transient.
To create a training dataset with rich dynamics, we discard all the time series that laminarize, i.e., the time series whose maximum kinetic energy is larger than the laminarization threshold, $k_l=0.48$, where $k_l$ is selected to be close to the (asymptotic) laminarization value, $k=0.5$.
Approximately $\simeq 28\%$ of all the time series are discarded in the ensemble due to laminarization. Hence, the dataset consists of 1436 time series, which are used to train, validate and test the echo state networks  (Section~\ref{sec:ESN}).
Specifically, we use ten time series for training and validating an ensemble of echo state networks. Each network is trained on the same ten times series and receives the entire state of the system with downsampled $\delta t = 4dt$ \cite{racca2022statistical}(more details in appendix \ref{sec:train}). We use an ensemble of ten networks due to the random component of the initialization of the matrices in echo state networks (section \ref{sec:ESN}). A summary of the parameters used in the generation of the dataset is shown in \cref{tab:dataset generation}.

\section{Echo State Networks}
\label{sec:ESN}

In time-series prediction, the data is sequentially ordered in time. 
Recurrent neural networks are designed to infer the correlation within sequential data through an internal hidden state, which is updated at each time step. Because of the long-lasting time dependencies of the hidden state, however, training RNNs with backpropagation through time is notoriously difficult \cite{werbos1990backpropagation}. Echo state networks overcome this problem by nonlinearly expanding the inputs into a high-dimensional system, the reservoir, which also acts as the memory of the system \cite{lukovsevivcius2012practical}. Hence, the output is computed as a linear combination of the reservoir's dynamics, whose weights are the only trainable parameters of the system. Thanks to this specific architecture, training echo state networks consists of a straightforward linear regression problem, which bypasses backpropagation through time. 

\begin{figure}[H]
\centering
\includegraphics[width=.5\textwidth]{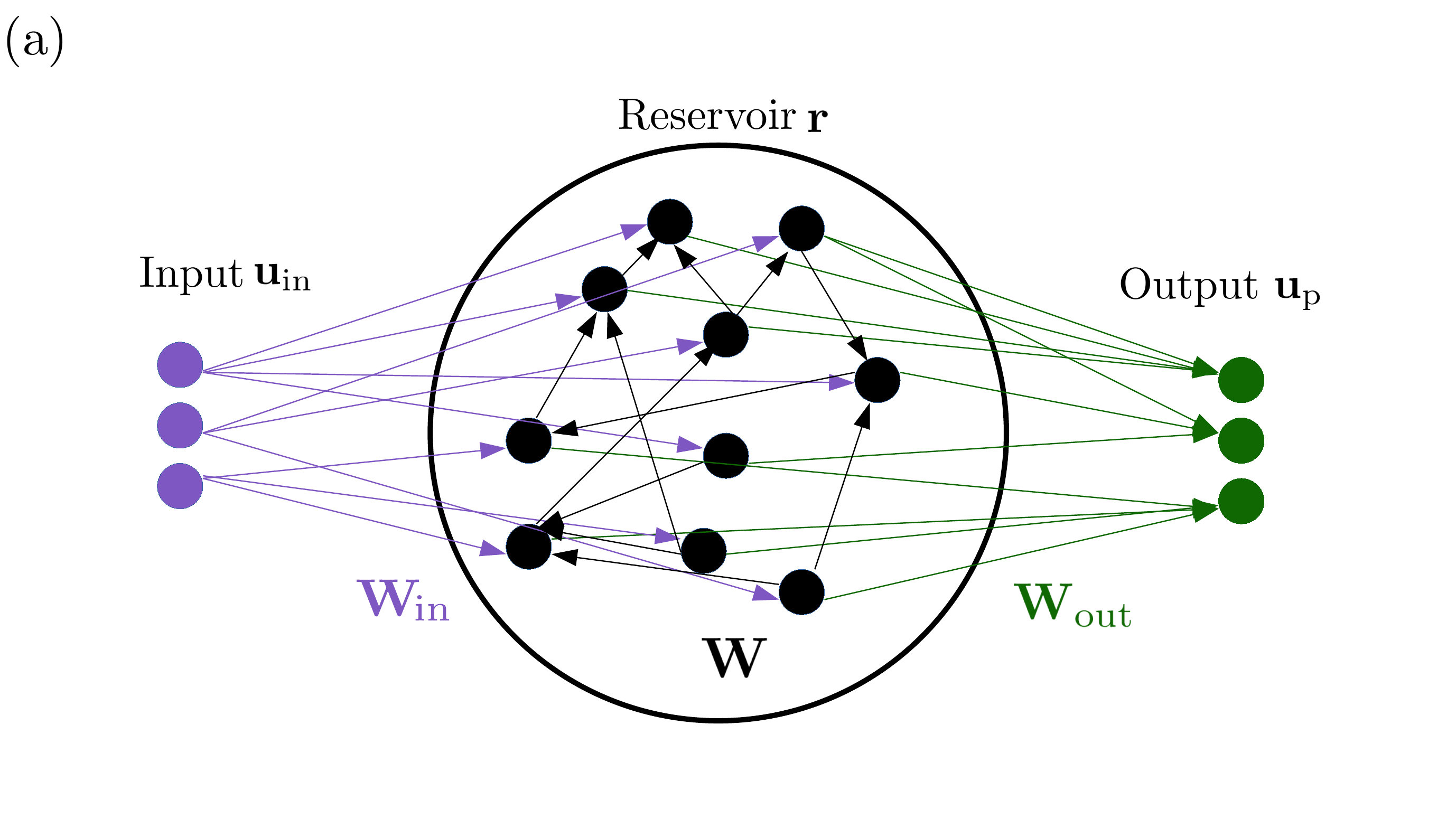}
\includegraphics[width=1.\textwidth]{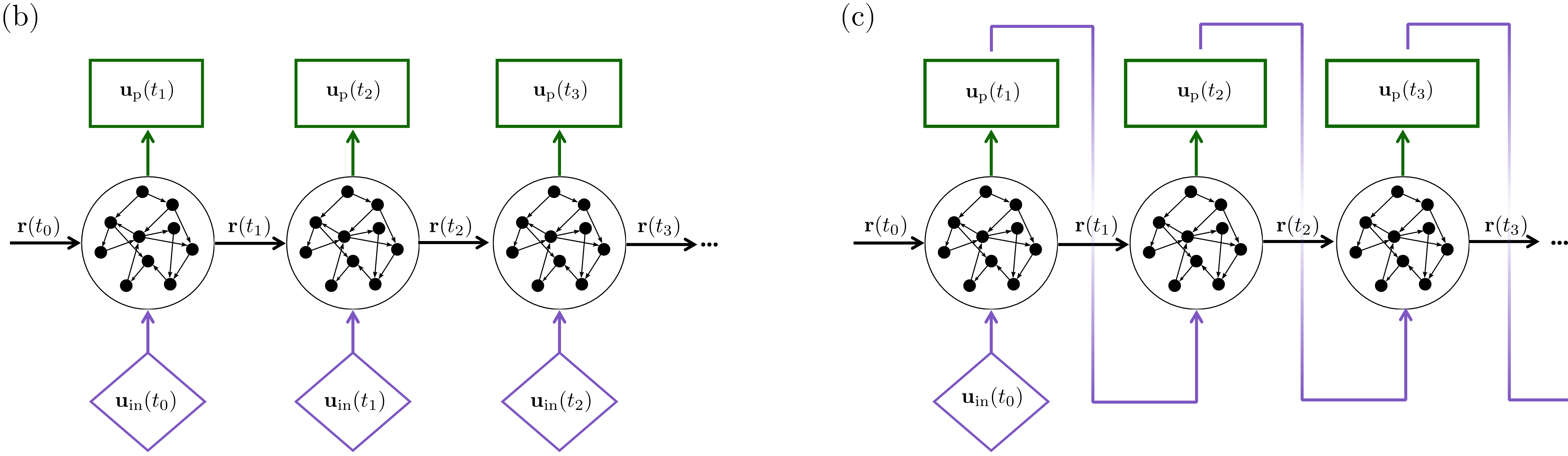}
\caption{(a) Schematic representation of the echo state network. (b) Open-loop and (b) closed-loop  configurations.}
\label{fig:ESN}
\end{figure}

As shown in \cref{fig:ESN}(a), in an echo state network, at any time $t_i$ the input vector, $\textbf{u}_{\mathrm{in}}(t_i) \in \mathbb{R}^{N_u}$, is mapped into the reservoir state, by the input matrix, $\mathbf{W}_{\mathrm{in}} \in \mathbb{R}^{N_r\times (N_u+1)}$, where $N_r \gg N_u$. 
The reservoir state, $\textbf{r} \in \mathbb{R}^{N_r}$, which is updated at each time iteration as a function of the current input and its previous value, is used to compute the predicted output,  $\textbf{u}_{\mathrm{p}}(t_{i+1})\in \mathbb{R}^{N_u}$. The reservoir state and output are governed by the discrete dynamical equations, respectively,

\begin{gather}
\label{state_step}
        \textbf{r}(t_{i+1}) = \textrm{tanh}\left(\mathbf{W}_{\mathrm{in}}[\mathbf{\hat{u}}_{\mathrm{in}}(t_i);b_\mathrm{in}]+\mathbf{W}\textbf{r}(t_i)\right), \\ \mathbf{u}_{\mathrm{p}}(t_{i+1}) = [\mathbf{r}(t_{i+1});1]^T\mathbf{W}_{\mathrm{out}};
\end{gather}
where $\hat{(\;\;)}$ indicates normalization by the range component-wise, $\mathbf{W} \in \mathbb{R}^{N_r\times N_r}$ is the state matrix, $b_{\mathrm{in}}$ is the input bias and 
$\mathbf{W}_{\mathrm{out}} \in \mathbb{R}^{(N_{r}+1)\times N_{u}}$ is the output matrix. 
Matrices 
$\mathbf{W}_{\mathrm{in}}$ and $\mathbf{W}$ are (pseudo)randomly generated and fixed, whilst the weights of the output matrix, $\mathbf{W}_{\mathrm{out}}$, are computed by training the network. The input matrix,
$\mathbf{W}_{\mathrm{in}}$, has only one element different from zero per row, which is sampled from a uniform distribution in $[-\sigma_{\mathrm{in}},\sigma_{\mathrm{in}}]$, where $\sigma_{\mathrm{in}}$ is the input scaling. The state matrix, $\textbf{W}$, is an Erdős-Renyi matrix with average connectivity $d$, in which each neuron (each row of $\mathbf{W}$) has on average only $d$ connections (non-zero elements), which are obtained by sampling from a uniform distribution in $[-1,1]$; the entire matrix is then rescaled by a multiplication factor to set the spectral radius, $\rho \leq 1$ to enforce the echo state property \cite{lukovsevivcius2012practical}. 
The bias in the inputs and outputs layers are added to break the inherent symmetry of the basic ESN architecture \cite{lu2017reservoir,huhn2020learning}.  The input bias, $b_{\mathrm{in}}$ is a hyperparameter, selected in order to have the same order of magnitude of the normalized inputs, $\mathbf{\hat{u}}_{\mathrm{in}}$, whilst  the output bias is determined by training the weights of the output matrix, $\mathbf{W}_{\mathrm{out}}$.

The ESN can be run either in open-loop or closed-loop. 
In the open-loop configuration (panel (b) in \cref{fig:ESN}), we feed the data as the input at each time step to compute and store the reservoir dynamics, $\mathbf{r}(t_i)$. 
In the initial transient of this process, which is the washout interval, we do not compute the output, $\textbf{u}_{\mathrm{p}}(t_i)$. 
The purpose of the washout interval is for the reservoir state to satisfy the echo state property. In doing so the reservoir state becomes (i) up-to-date with respect to the current state of the system and (ii) independent of the arbitrarily chosen initial condition, $\textbf{r}(t_0) = {0}$. 
After washout, we train the output matrix, $\mathbf{W}_{\mathrm{out}}$, by minimizing the mean square error between the outputs and the data over the training set.
Training the network on $N_{\mathrm{tr}} + 1$ points  needs neither backpropagation nor gradient descent, but it only needs a solution of a linear system (ridge regression) 
\begin{equation}
\label{RidgeReg}
    (\mathbf{R}\mathbf{R}^T + \beta \mathbf{I})\mathbf{W}_{\mathrm{out}} = \mathbf{R} \mathbf{U}_{\mathrm{d}}^T,
\end{equation}
\noindent where $\mathbf{R}\in\mathbb{R}^{(N_r+1)\times N_{\mathrm{tr}}}$ and $\mathbf{U}_{\mathrm{d}}\in\mathbb{R}^{N_u\times N_{\mathrm{tr}}}$ are the horizontal concatenation of the reservoir states with bias, $[\mathbf{r};1]$, and of the output data, respectively; 
$\mathbf{I}$ is the identity matrix and $\beta$ is the Tikhonov regularization parameter \cite{tikhonov2013numerical}. 

In the closed-loop configuration (panel (c) in \cref{fig:ESN}), starting from an initial data point as an input and an initial reservoir state obtained after the washout interval, the output, $\textbf{u}_{\mathrm{p}}$, is fed back to the network as an input for the next time step prediction. 
In doing so, the network is able to autonomously evolve in the future. 
The closed-loop configuration is used during validation and test, but not during training.

\subsection{Validation}

\label{sec:val}

During validation, we use part of the data to select the hyperparameters of the network by minimizing the logarithm of the mean square error of the kinetic energy with respect to the data (section \ref{sec:MFE}). 
We use the logarithm because the error varies by multiple orders of magnitude with different hyperparameters.
In this work, we tune the input scaling, $\sigma_{\mathrm{in}}$, spectral radius, $\rho$, and Tikhonov parameter, $\beta$, which are the
key hyperparameters for the performance of the network \cite{lukovsevivcius2012practical,racca2021robust}.  
We use Bayesian Optimization to select $\sigma_{\mathrm{in}}$ and $\rho$, and perform a grid search within  $[\sigma_{\mathrm{in}},\rho]$ to select $\beta$ \cite{racca2021robust}. We explore the hyperparameter space $[0.1,10]\times[0.1,1]$ for $[\sigma_{\mathrm{in}},\rho]$ in logarithmic scale and the grid $[10^{-6}, 10^{-9},10^{-12}]$ for $\beta$. The Bayesian Optimization starts from a grid of $4\times4$ points in the $[\sigma_{\mathrm{in}},\rho]$ domain, and then it selects nine additional points through the gp-hedge algorithm \cite{hoffman2011portfolio}. We set $b_{\mathrm{in}}=0.1$, $d=20$ and add gaussian noise with zero mean and standard deviation, $\sigma_n=0.01\sigma_u$, where $\sigma_u$ is the standard deviation of the data component-wise, to the training and validation data \cite{vlachas2020backpropagation}. Adding noise to the data improves the performance of ESNs in chaotic dynamics because the network explores the region around the attractor and becomes more robust to errors when forecasting the future evolution of the trajectory \cite{vlachas2020backpropagation,lukovsevivcius2012practical}. A summary of the hyperparameters is shown in \cref{tab:ESN hyperparameters}

\begin{table}[H]
\centering
\caption{Echo state networks' hyperparameters. Multiple values indicate that the parameter is optimized in the shown range.}
\begin{tabular}{l l l}
\toprule
Parameter $\qquad$ &  & Value \\
\midrule
$\rho$ & Spectral radius  & $[0.1,1]$ \\
$\sigma_{\mathrm{in}}$ & Input scaling & $[0.1,10]$ \\
$\beta$ & Tikhonov parameter & $[10^{-6}, 10^{-9},10^{-12}]$  \\
$d$ & Connectivity & $20$  \\
$b_{\mathrm{in}}$ & Input bias & $0.1$  \\
$\sigma_n$ & Standard deviation of the noise in the training inputs $\qquad$ & $0.01\sigma_{u}$  \\
\toprule
\end{tabular}
\label{tab:ESN hyperparameters}
\end{table}

\begin{figure}[H]
\centering
\includegraphics[width=.9\textwidth]{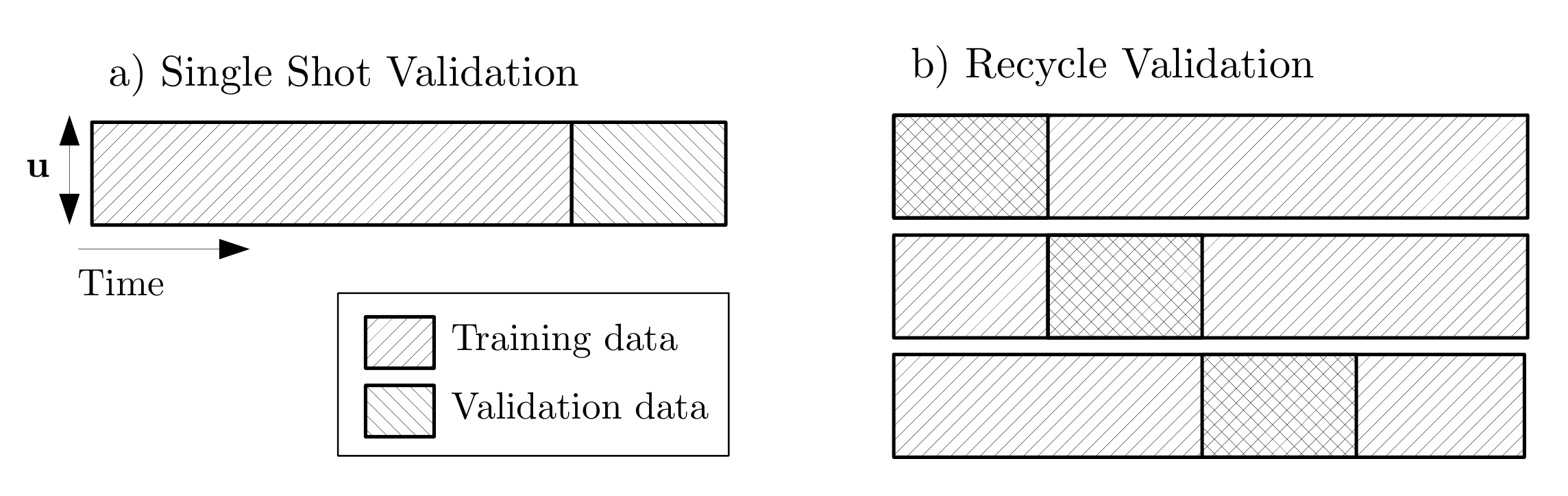}
\caption{Schematic representation of the validation strategies, where $\mathbf{u}$ represents the number of degrees of freedom of the data at a single timestep. Three validation intervals are shown for the Recycle Validation \cite{racca2021robust}.}
\label{fig:Vals}
\end{figure}

To select the hyperparmeters, we use Single Shot Validation (SSV) and Recycle Validation (RV) \cite{racca2021robust}. The Single Shot Validation is the most commonly used validation strategy for RNNs in the literature \cite{lukovsevivcius2019efficient}. It consists of splitting the data in a training set and a single small validation set subsequent in time to the training set (panel (a) in  \cref{fig:Vals}). The size of the validation set is limited by the chaotic nature of the signal, which causes the predicted trajectory to diverge exponentially in time from the data (section \ref{sec:lyap}). Because the hyperparmeters are tuned with respect to a small interval, which is not representative of the available dataset, the validated networks may perform poorly in the test set \cite{racca2021robust}. To improve the selection of hyperparameters, the K-Fold Cross Validation (KCV) \cite{stone1974cross}, which validates the network in multiple intervals of the dataset, is sometimes used. The K-Fold Cross Validation is computationally expensive, because the network needs to be trained multiple times for each set of hyperparameters. To overcome the high computational cost of the KCV, whilst keeping the improved performance, the authors of \cite{racca2021robust} proposed the Recycle Validation (RV). In the RV, the network is trained only once on the entire dataset and validation is performed on multiple intervals already used for training. This is possible because recurrent neural networks operate in two configurations (open-loop and closed-loop), which means that the networks can be validated in closed-loop on data used for training in open-loop (panel (b) in  \cref{fig:Vals}).

\section{Time-accurate prediction of extreme events}

The time-accurate prediction of extreme events is the  prediction in the near future of the occurrence of an extreme event given data at the current time. Forecasting an incoming extreme event can be viewed as a {\it binary  classification} problem, in which we evaluate whether (or not) the event will happen. Because extreme events are also rare, the classification operates on an imbalanced dataset, in which the occurrence of an extreme event is significantly less likely than the non-occurrence of an event. To assess the capability of echo state networks to time-accurately predict extreme events, therefore, we tailor figures of merit that take into account the imbalance nature of the dataset \cite{he2009learning}. In this section, we (i) assess the performance of the network to accurately predict extreme events from multiple points along the dataset through precision, recall and the F-score; (ii) compute how far in advance we can foretell extreme events through the prediction horizon; and (iii) extend the  analysis to multiple Reynolds number to verify the robustness of the networks and results. 
\label{sec:time-acc}
%
\subsection{Precision and recall}
\label{sec:prec}

\begin{figure}[H]
\centering
\includegraphics[width=1.\textwidth]{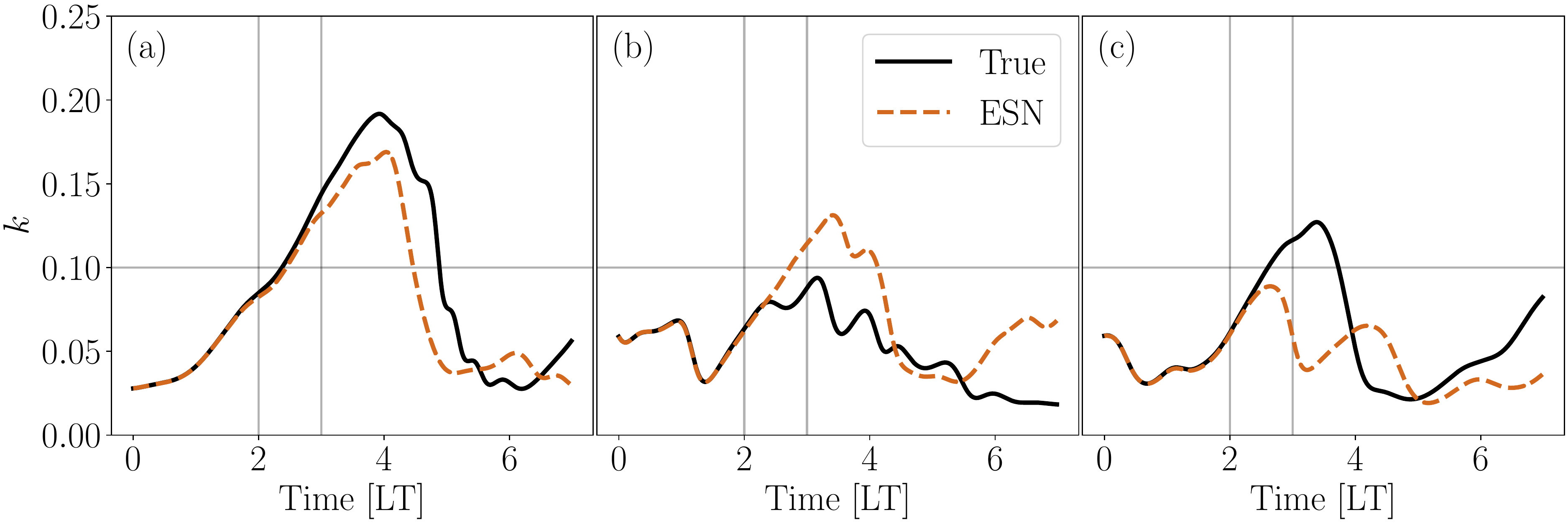}
\caption{(a) True positive, (b) false positive, and (c) false negative predictions of an extreme event with prediction time equal to 2LTs. The horizontal line indicates the extreme event threshold, $k_e$. The vertical lines indicate the time window of the prediction.}
\label{fig:prediction}
\end{figure}

\begin{figure}[H]
\centering
\includegraphics[width=1.\textwidth]{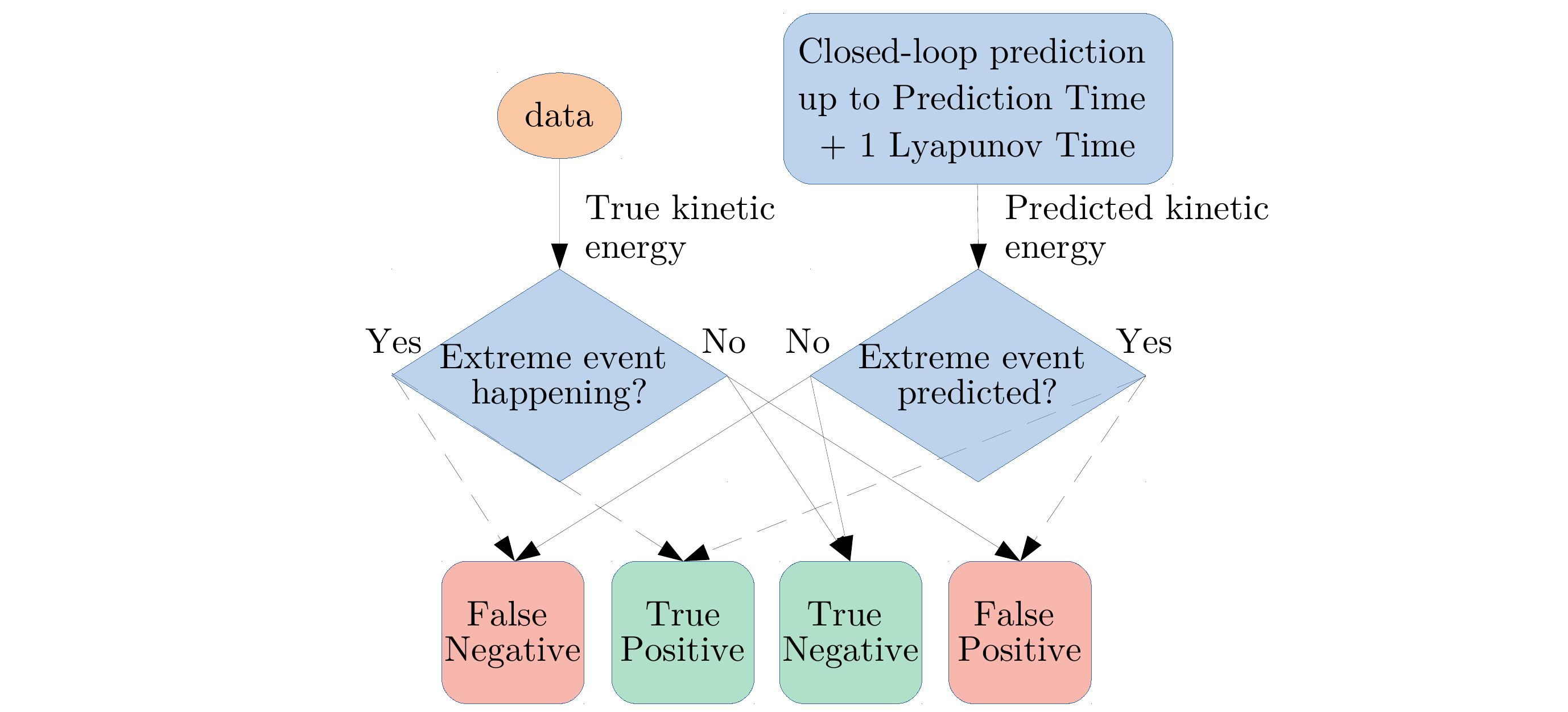}
\caption{Schematic representation of the prediction of extreme events with echo state networks. We evaluate whether the predicted kinetic energy and the true kinetic energy display an extreme event in the interval lasting 1 Lyapunov time after the prediction time. A more detailed block diagram can be found in the supplementary material.}
\label{fig:pr_scheme}
\end{figure}

We assess the capability of the networks to predict extreme events.
To do so, we monitor the system in real-time by letting the network evolve autonomously (closed-loop) to forecast the future behaviour of the flow  in the test set. We define the Prediction Time (PT) to evaluate if the network is predicting an extreme event in a specific window in the future. For a user-defined prediction time, we evaluate if the prediction generated by the network crosses the extreme event threshold, $k_e$, in the 1LT interval subsequent to the prediction time. The prediction time determines how far ahead into the future we are assessing the forecast of the network. For example, with a PT $=$ 2LTs, we evaluate if the network is forecasting an event between 2LTs and 3LTs in the future. 
There are three cases we consider: 
(i) if the network's prediction crosses $k_e$ and the true signal crosses $k_e$, we have a true positive;
(ii) if the network's signal crosses $k_e$ and the true signal does not, we have a false positive; and 
(ii) if the network's signal does not cross $k_e$ and the true signal crosses $k_e$, we have a false negative (\cref{fig:prediction} and \cref{fig:pr_scheme}). 

To evaluate the performance of the networks to forecast extreme events, we use precision, p, recall, r, and the F-score, F, which are defined as \cite{he2009learning}
\begin{equation}
\label{precision-recall}
    \textrm{p} = \frac{\textrm{TP}}{\textrm{TP}+\textrm{FP}}, \qquad \textrm{r} = \frac{\textrm{TP}}{\textrm{TP}+\textrm{FN}}, \qquad
    \textrm{F} = \frac{2}{\mathrm{p}^{-1} + \mathrm{r}^{-1}},
\end{equation}
where TP, FP and FN are true positives, false positives and true negatives, respectively.
Precision measures how many of the predicted events actually happen; 
recall measures how many events are predicted out of the total number of events,  which gives an indication on how many events are missed out by the prediction; and the F-score is the geometric average of the two. Therefore, p$=1$ means that all the predicted events happened, r$=1$ means that all the events have been predicted, and F$=1$ means that all the events were predicted and all predictions were correct.
\vfill
\textcolor{white}{a}

\begin{figure}[H]
\centering
\includegraphics[width=1.\textwidth]{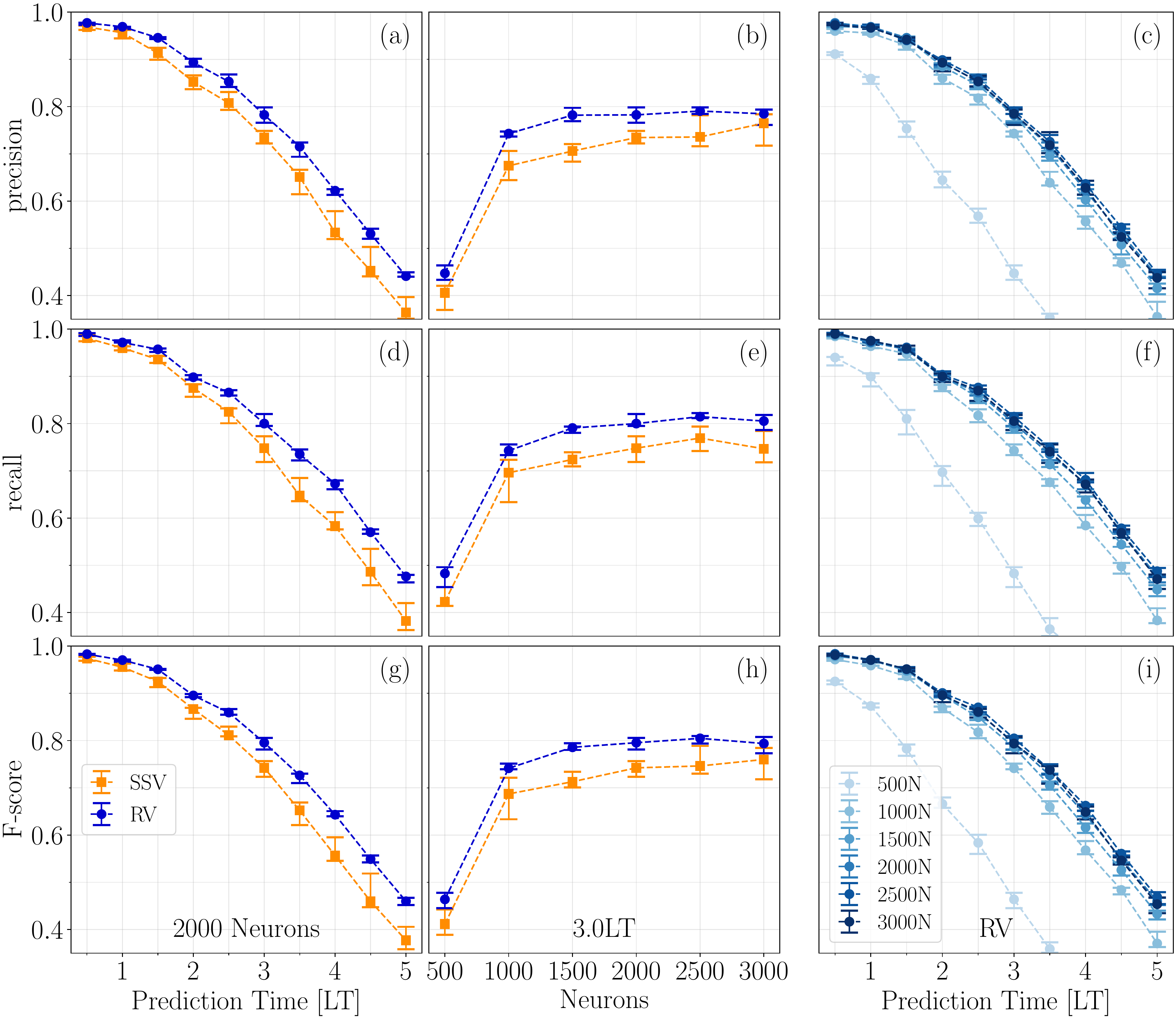}
\caption{25$th$, 50$th$ and 75$th$ percentiles of precision, recall, and F-score, as a function of the prediction time and number of neurons for the Single Shot Validation (SSV), and Recycle Validation (RV). (a)-(d)-(g) have fixed number of neurons, (b)-(e)-(h) have a fixed prediction time of $3$LTs, and (c)-(f)-(i) are for the RV only.}
\label{fig:F}
\end{figure}

\cref{fig:F} shows the results obtained by predicting the occurrence of extreme events for different prediction times in the test set. The results are obtained from 9000 starting points, which span 150 time series and consist of 360 extreme events. We select the starting points to be spaced 1 Lyapunov times from each other. 
First, the Recycle Validation outperforms the Single Shot Validation in all the figures of merit, which implies that networks optimized with RV correctly predict more events than networks optimized with SSV. In large reservoirs (neurons $\geq1500$), the Recycle Validation improves the F-score with respect to the Single Shot Validation by an amount comparable to the improvement obtained by increasing the size of the reservoir by hundreds of neurons. For example, the 1500-neurons networks trained with the Recycle Validation have a  median of the F-score of $\mathrm{F_{RV}}=0.79$, which is greater than the median of the F-score of the 3000-neurons networks trained with Single Shot Validation ($\mathrm{F_{SSV}}=0.76$) (panel (h)).
Second, the networks are able to accurately predict extreme events multiple Lyapunov times in advance, with $\mathrm{F}\geq0.9$ for 2LTs prediction time and up to $\mathrm{F}=0.8$ for 3LTs prediction time (panel (g)). This means that we predict almost all the events $(\simeq 90\%)$ (panel (d)) with only a few false positives $(\simeq 10\%)$ (panel (a)) 2LTs before the events happen. 
Third, panels (c)-(f)-(i) display the performance for the Recycle Validation as the reservoir increases. After a significant improvement in the performance between 500 and 1000 neurons, the performance of the networks remains approximately constant for networks with more than 1500 neurons. This indicates that 1500 neurons are able to learn the  chaotic dynamics needed to forecast the evolution of the MFE system. Crucially, the marked difference in performance between different validation strategies (panels (b)-(e)-(h)) means that for large reservoirs tuning the hyperparameters is more effective than increasing the number of neurons. 

\subsection{Prediction horizon of extreme events}

\label{sec:Phee}

To assess how far in advance the networks predict extreme events, we use the prediction horizon for extreme events, $\mathrm{PH}_{\mathrm{ee}}$. The prediction horizon is a metric to evaluate the time-accurate prediction of chaotic dynamics with data-driven methods \cite{pathak2018hybrid,racca2021robust,doan2021short, vlachas2020backpropagation}. In this work, it is defined as the time interval during which the absolute error between the kinetic energy predicted by the ESN, $k_{\mathrm{ESN}}$, and the true kinetic energy, $k_{\mathrm{True}}$ is bounded as  
\begin{equation}
    \frac{|k_{\mathrm{ESN}}(t) - k_{\mathrm{True}}(t)|}{k_e - \overline{k}} < 0.2,
    \label{eq:PH}
\end{equation}
where $\overline{k}$ is the time-average of the kinetic energy, $k_e$ is the extreme event threshold, $k_e - \overline{k}$ acts as a normalization factor, and 0.2 is a user-defined threshold.  

\begin{figure}[H]
\centering
\includegraphics[width=1.\textwidth]{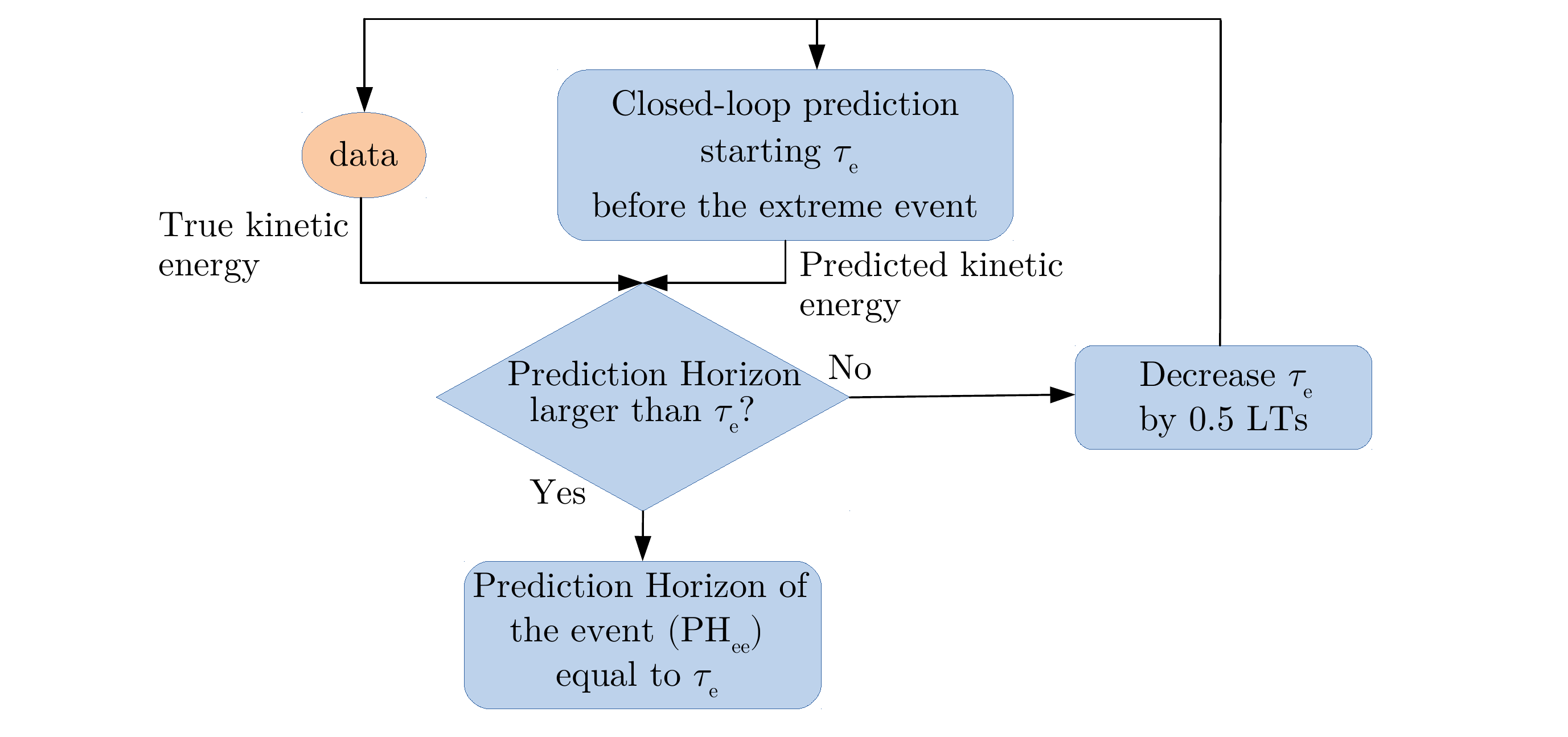}
\caption{Block diagram for the computation of the prediction horizon for extreme events, $\mathrm{PH}_{\mathrm{ee}}$. We evaluate the prediction horizon \eqref{eq:PH} of the closed-loop prediction starting from $\tau_e$ before the event. If the prediction horizon is larger than $\tau_e$, $\mathrm{PH}_{\mathrm{ee}}=\tau_e$, if not, we decrease $\tau_e$ and evaluate the prediction horizon starting from the new $\tau_e$ before the event. The time $\tau_e$ is initialized to 10 LTs. A more detailed diagram is shown in the supplementary material.}
\label{fig:phee_scheme}
\end{figure}

\begin{figure}[H]
\centering
\includegraphics[width=1.\textwidth]{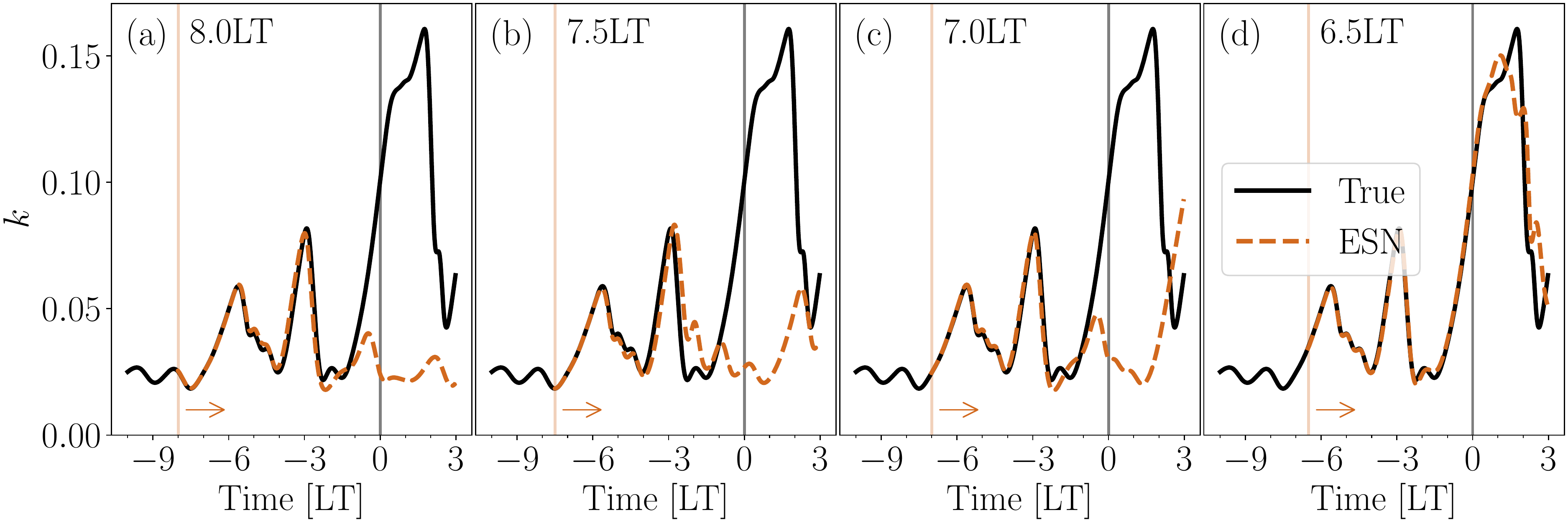}
\caption{Last four evaluations before the event is predicted with $\mathrm{PH}_{\mathrm{ee}}$=6.5 LTs. In each panel the prediction starts from 0.5LTs closer to the extreme event with respect to the previous panel.}
\label{fig:PHEE_plot}
\end{figure}

We define the prediction horizon for extreme events, $\mathrm{PH}_{\mathrm{ee}}$, by identifying the instants $t^{(e)}$ at which extreme events start, such that $k(t) = k_e$ and $dk(t)/dt>0$ at $t=t^{(e)}$. The value of $t^{(e)}$ is obtained by prepocessing the test set data. 
For the $i$-th event,
we start a closed-loop prediction of the ESN at time $t^{(e)}_i - \tau_e$, $\tau_e=10$LTs, and evaluate whether the prediction horizon is larger than $\tau_e$. If the prediction horizon is larger than $ \tau_e$, then $\mathrm{PH}_{\mathrm{ee}}$ is equal to $\tau_e$, if the prediction horizon is smaller than $\tau_e$, we decrease $\tau_e$ by 0.5LTs, evaluate whether the prediction horizon is larger than $\tau_e$, and so on and forth (\cref{fig:phee_scheme,fig:PHEE_plot}).

\begin{figure}[H]
\centering
\includegraphics[width=.5\textwidth]{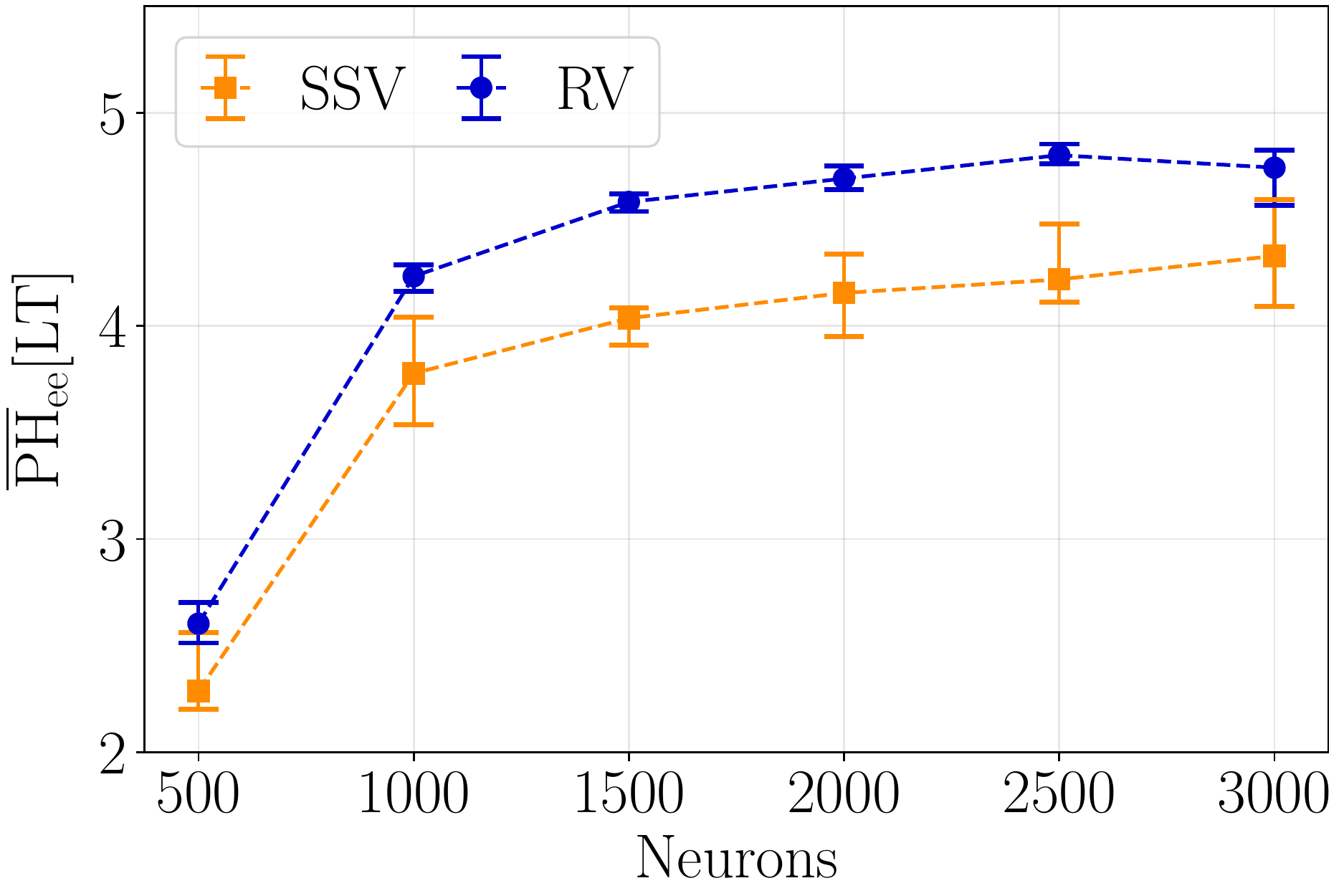}
\caption{25$th$, 50$th$ and 75$th$ percentiles for the average prediction horizon of extreme events as a function of the number of neurons for the Single Shot Validation (SSV) and the Recycle Validation (RV).}
\label{fig:PHEE}
\end{figure}

\cref{fig:PHEE} shows the average $\mathrm{PH}_{\mathrm{ee}}$ over 500 extreme events for different sizes of the reservoirs. The networks are able to predict the extreme events up to almost 5LTs in advance on average. The performance increases sharply in smaller networks, and levels off for larger networks. The Recycle Validation outperforms the Single Shot Validation in all the figures of merit, and in most cases generates an improvement in performance comparable to the increase in size of the reservoir of hundred, and even thousands, of neurons. For example, the 1000-neurons networks trained with Recycle Validation predict $50\%$ of the events (median of $\overline{\mathrm{PH}}_{\mathrm{ee-RVC}}=4.23 \mathrm{LTs}$) with approximately the same prediction horizon of the 2500-neurons networks trained with Single Shot Validation (median of $\overline{\mathrm{PH}}_{\mathrm{ee-SSV}}=4.22\mathrm{LTs}$). 


\subsection{Analysis at different Reynolds numbers}

In order to further corroborate the results of section \ref{sec:prec}-\ref{sec:Phee}, we analyse the time-accurate prediction of the MFE system's extreme events for different Reynolds numbers, Re=[600,800,1000], for which the Lyapunov times are LT=[$0.0135^{-1}$, $0.0126^{-1}$, $0.0121^{-1}$], respectively. We study this enlarged range of Reynolds numbers to verify the robustness of echo state networks in time-accurately predict chaotic dynamics with respect to changes the physical parameter, which, here, is the Reynolds number. For each Reynolds number, the networks are retrained and validated (section \ref{sec:MFE} and appendix \ref{sec:train}).  The flow shows qualitatively similar dynamics in the new range of Reynolds numbers: a chaotic transient with extreme events that eventually converges to the fixed point $a_1=1,a_2=\dots=a_9=0$. 

\begin{figure}[H]
\centering
\includegraphics[width=.9\textwidth]{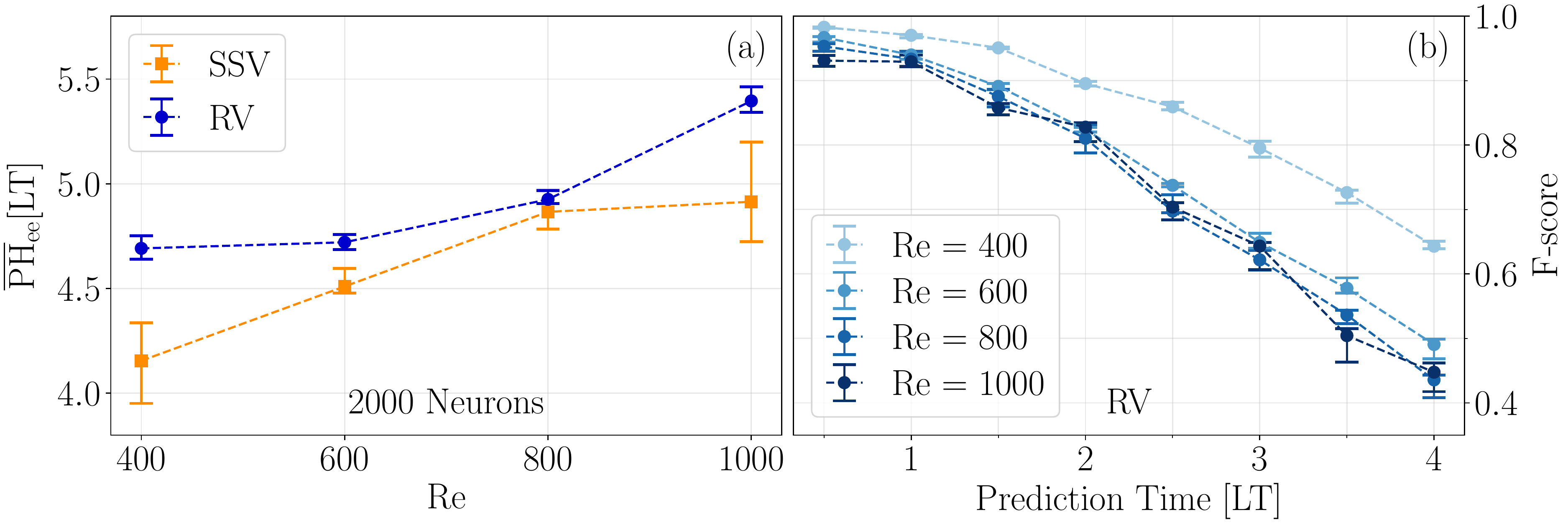}
\caption{25$th$, 50$th$ and 75$th$ percentiles for the (a) average prediction horizon of extreme events and (b) F-score as a function of the Reynolds Number for 2000 neurons networks.}
\label{fig:Reys}
\end{figure}

\cref{fig:Reys} shows the results for the prediction horizon for extreme events and the F-score. 
On the one hand, increasing the Reynolds number increases the prediction horizon for extreme events (panel (a)). On the other hand, the F-score deteriorates from $\mathrm{Re}=400$ to the other cases, which perform similarly one to the other (panel (b)).
In agreement with the observed behaviour in the $\mathrm{Re}=400$ case, the Recycle Validation outperforms the Single Shot Validation in all figures of merit in the new range of Reynolds values. These results indicate that the analysis for the $\mathrm{Re}=400$ case extends to a broader range of Reynolds numbers, which means that ESNs are robust machines with respect to changes in the physical parameters of the system.

\section{Statistical prediction of extreme events}

\label{sec:stats}

In section \ref{sec:time-acc}, we analyse time-accurate predictions. In this section, we move our focus on to the flow long-term statistics. In ergodic systems, the flow statistics are robust quantities, which are not significantly affected by the butterfly effect, in contrast to the time-accurate prediction\footnote{In chaotic flows, the trajectories cannot be time-accurately predicted indefinitely because the tangent space of chaotic flows is exponentially unstable (butterfly effect). This is an intrinsic limit to which the time-accurate prediction of chaos is subject (see section \ref{sec:lyap}). }. 
Physically, nearby trajectories evolve within the same bounded region of the phase space, which means that they share the same long-term statistics.

We analyse the capability of the networks to predict the statistical behaviour of the flow through long-term predictions. Long-term predictions are closed-loop predictions that last several tens of Lyapunov times and diverge from the true data. The long term predictions are generated from 500 different starting points in the training set, from which we generate a total of 500 different time series per network (one per starting point). Each time series is obtained by letting the ESN evolve in closed-loop for 4000 time units ($\simeq 65$LTs). We discard all the time series that laminarize. We compute the statistics produced by the ESN using the remaining time series. This procedure is analogous to the generation of the data through the numerical integration of the governing equations discussed in section \ref{sec:MFE}.
We use the long-term prediction to (i) assess the prediction of the Probability Density Function (PDF) of the kinetic energy for different sizes of the reservoir; and (ii) predict the statistics of flow variables in the physical domain $L_x\times L_y\times L_z$. In all cases, we compare the statistical prediction of the networks with the statistics of the training set. We do so because the objective of predicting the statistics is to improve our knowledge via the networks from the already-available knowledge in the training data. This is a markedly different approach from previous works such as \citet{snrinivasan2020}, where the large datasets available during training already have converged statistics, and the networks are trained to replicate the statistics, but not to extrapolate them.
In this study, we use small training datasets with non-converged statistics to train the network and extrapolate the statistics. This  improves the statistical knowledge of the flow with respect to the available data. 

To assess the ESN performance on the prediction of the statistics of the kinetic energy, we use the Kantorovich metric and the mean logarithmic error, which are computed with respect to the true Probability Density Function of the kinetic energy, $\mathrm{PDF}_{\mathrm{True}}(k)$. The first metric is the first-order Kantorovich Metric \cite{kantorovich1942translocation}, also known as the Wasserstein distance or Earth mover's distance \cite{monge1781memoire}, measures the distance between two PDFs. Intuitively, it evaluates the minimum work to transform one PDF to  another. It is computed as
\begin{equation}
    \mathcal{K} = \int_{-\infty}^{\infty}|\mathrm{CDF}_{\mathrm{True}}(k) - \mathrm{CDF}_j(k)|dk,
\end{equation}
where $\mathrm{CDF}$ is the Cumulative Distribution Function and $j$ indicates the PDF that we are comparing with the true data. We use the Kantorovich metric to assess the accuracy on the overall prediction of the PDF of the kinetic energy. If $\mathcal{K}$ is small, the two PDFs are similar to each other, while if $\mathcal{K}$ is large, the two PDFs significantly differ one from another.
The second metric is the mean logarithmic error, which is defined as 
\begin{equation}
    \mathrm{MLE} = \sum_{i=1}^{n_{b}}n_b^{-1} |\log_{10}(\mathrm{PDF}_{\mathrm{True}}(k)_i - \log_{10}(\mathrm{PDF}_j(k)_i)|,
\end{equation}
where the sum is performed over the total number of bins, $n_b$, used in the PDF. In case a bin has a value equal to zero and the logarithm is undefined, we saturate the logarithmic error in the bin to 1. We use the MLE to assess the prediction's accuracy on the tail of the PDF because the logarithm amplifies the errors in the small values of the tail.
\begin{figure}[H]
\centering
\includegraphics[width=.66\textwidth]{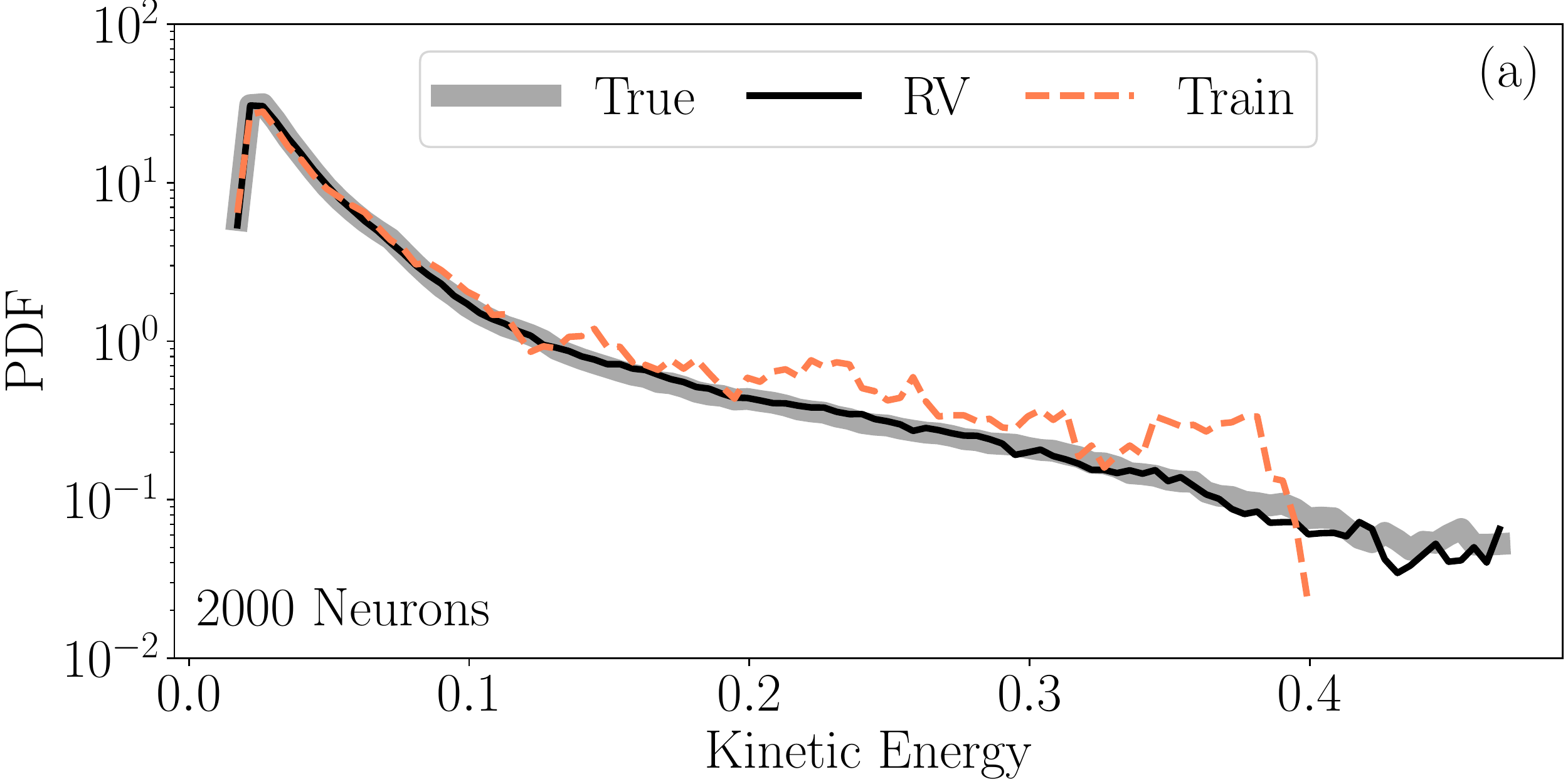}
\includegraphics[width=1.\textwidth]{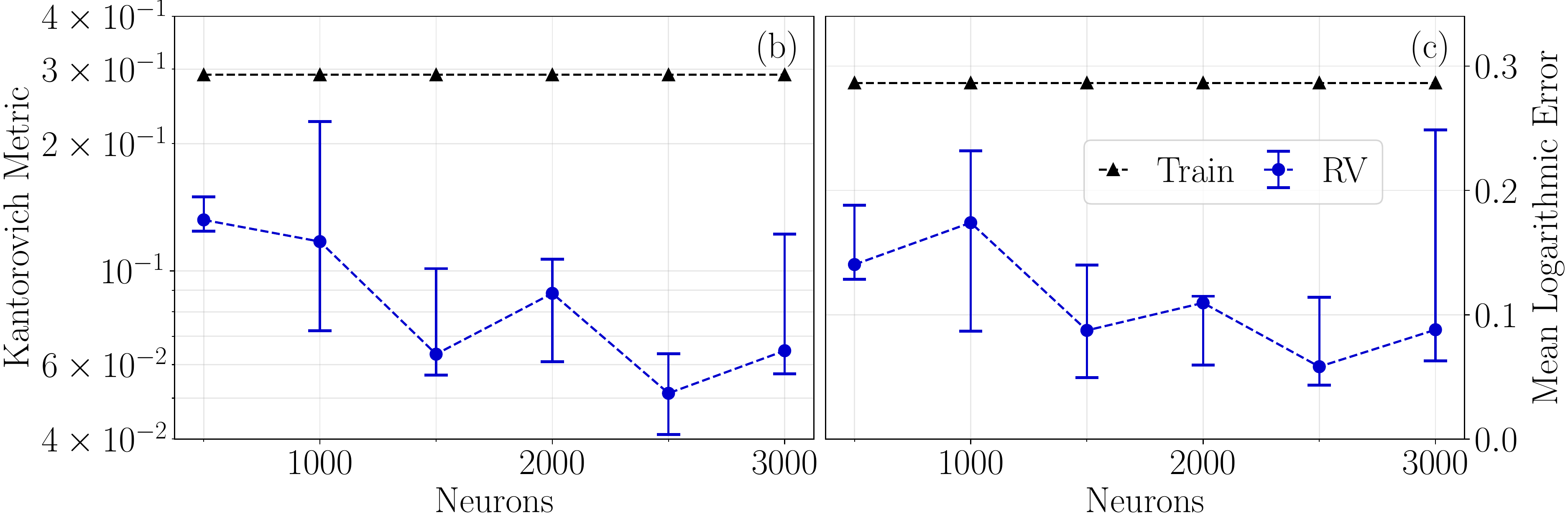}
\caption{(a) Probability density function (PDF) of the kinetic energy for the entire time series (True), training set (Train) and best performing 2000 neurons ESN for the Recycle Validation (RV). (b)-(c) 25$th$, 50$th$ and 75$th$ percentiles of (b) the Kantorovich metric and (c) the mean logarithmic error for the ESNs ensemble and training set.}
\label{fig:PDF}
\end{figure}
\cref{fig:PDF} shows the statistics of the kinetic energy for the ESNs optimized with the Recycle Validation and the training set.  Panel (a) shows the true PDF, the PDF predicted by the ESN and the training set PDF. Although the statistics of the training set is not close to the true statistics, the network learns the tail of the distribution by extrapolation. In other words, the echo state networks have learned a good approximation of the converged statistics from  a relatively short training set. 
Panels (b)-(c) show the Kantorovich metric and the mean logarithmic error for an ensemble of ten networks. On the one hand, the networks improve the statistics with respect to the training set in all cases analysed. On the other hand, there is no significant improvement in the performance of the networks when increasing their size. This means that the networks are able to both learn from an imperfect dataset and improve the prediction of the overall dynamics of the flow, even in small reservoirs. These results extend the findings of \citet{racca2022statistical} to smaller sizes of the reservoir.

\begin{figure}[H]
\centering
\includegraphics[width=1.\textwidth]{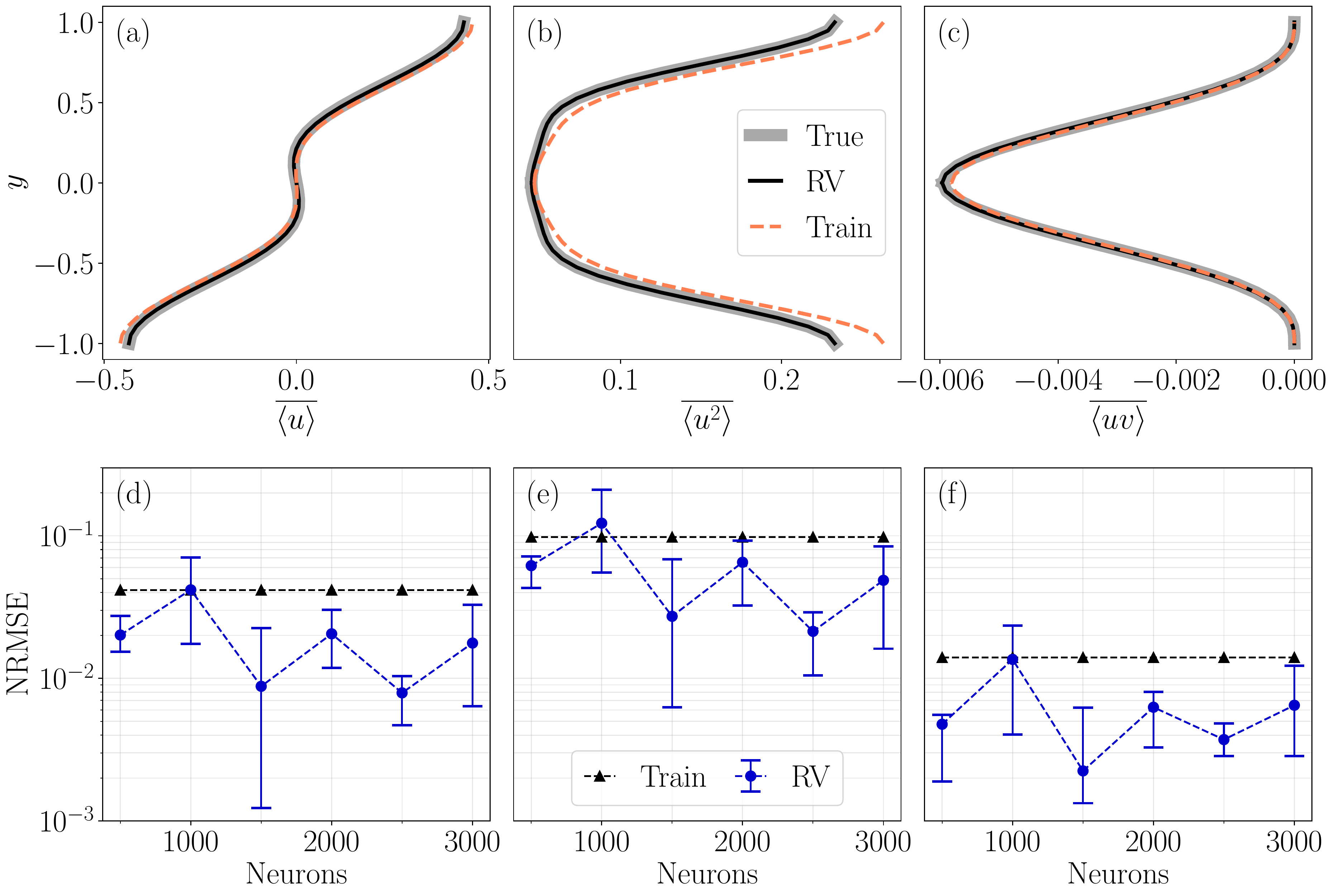}
\caption{(a) Profile of the spatial and time average of the streamwise velocity, (b) its square, and (c) Reynolds stress. ESNs profiles are the best performing networks with 2000 neurons. (d-f) 25$th$, 50$th$ and 75$th$ percentiles of the NRMSE in the variables from (a-c) for the networks ensemble and training set.}
\label{fig:v_stats}
\end{figure}

\begin{figure}[H]
\centering
\includegraphics[width=1.\textwidth]{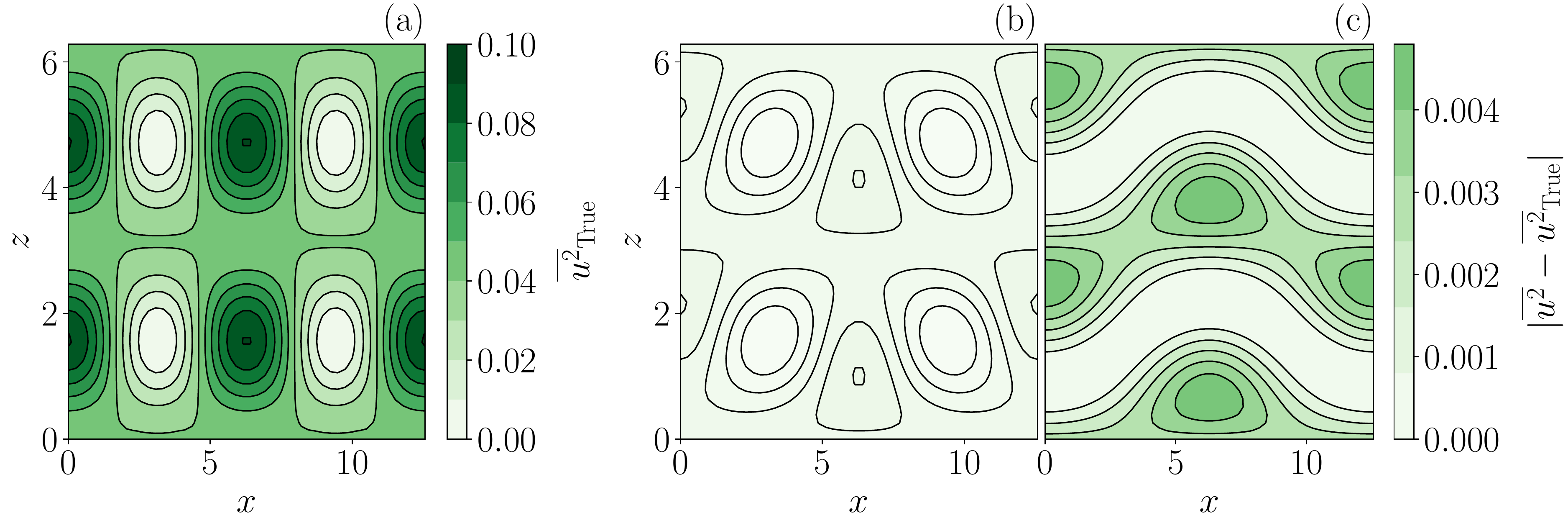}
\caption{(a) Time average of the square of the midplane velocity along $x$.  Error with respect to the true data for (b) a 2000 neurons network, and for (c) the training data.}
\label{fig:v_stats_flows}
\end{figure}

Finally, we study the statistics of the flow variables, following \citet{snrinivasan2020} and \citet{doan2021short}. We assess the prediction for the generic variable $\alpha$ through the Normalized Root Mean Square Error (NRMSE) with respect to the true data,
\begin{equation}
    \mathrm{NRMSE}(\alpha) = \frac{\sqrt{\frac{1}{N}\sum_{i}^{N}(\alpha_{i}-\alpha_{\mathrm{True}_{i}})^2}}{ \max(\alpha_{\mathrm{True}})   -  \min(\alpha_{\mathrm{True}})},
\end{equation}
where $( \allowbreak \max(\alpha_{\mathrm{True}})  \allowbreak - \min(\alpha_{\mathrm{True}}))$ is the range of $\alpha_{\mathrm{True}}$. 
Panels (d-f) of  \cref{fig:v_stats} show the NRMSE for the time, $\overline{(\;)}$, and spatial (along $x$ and $z$), $\langle (\;)\rangle $, average of the streamwise velocity, $u$, its square, $u^2$, and Reynolds stress, $uv$, whose profiles are plotted in panels (a-c). 
As seen for the PDF of the kinetic energy, (i) the ESN improves the statistics with respect to the available data and (ii) networks of different sizes perform similarly. Overall, there is a smaller improvement in the statistics of the flowfield than in the statistics of the kinetic energy of  \cref{fig:PDF}. This is because the networks are validated with a focus on the kinetic energy (section \ref{sec:val}). The time-average of the square of the velocity along $x$ in the midplane is shown in panel (a) of  \cref{fig:v_stats_flows} for the entire data. As seen for the spatial averages of  \cref{fig:v_stats}, the error for a network with 2000 neurons (panel (b)) is smaller than the error of the training set (panel (c)). These results show that echo state networks improve the prediction of the statistics of the flowfield in the physical space with respect to the training data. 
\vfill
\textcolor{white}{a}

\section{Control of extreme events}
\label{sec:control}
\begin{figure}[H]
\centering
\includegraphics[width=1.\textwidth]{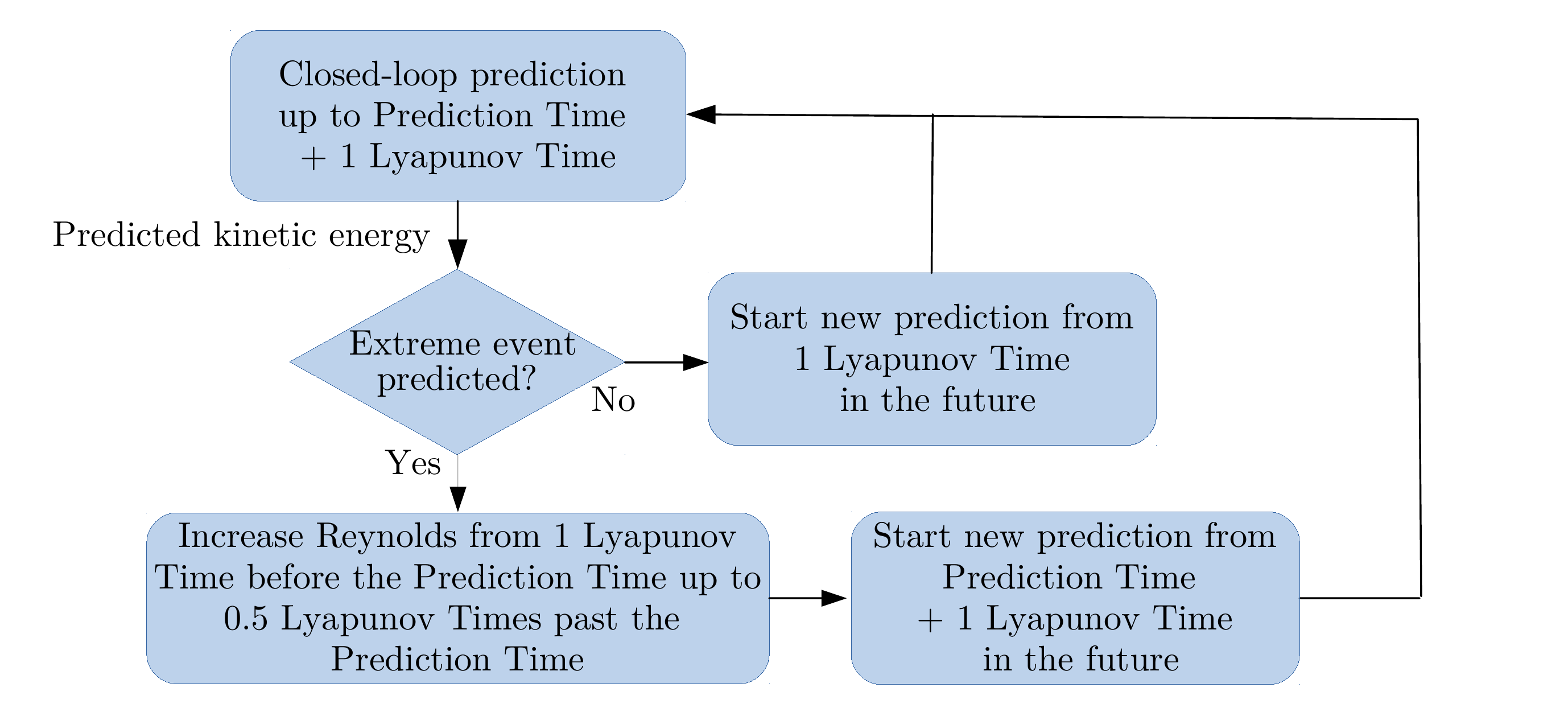}
\caption{Schematic representation of the control strategy. If the network predicts an event in the 1LT interval after the prediction time, we increase the Reynolds number before the event happens in order to suppress it. A more detailed schematic is shown in the supplementary material.}
\label{fig:control_scheme}
\end{figure}
\begin{figure}[H]
\centering
\includegraphics[width=1.\textwidth]{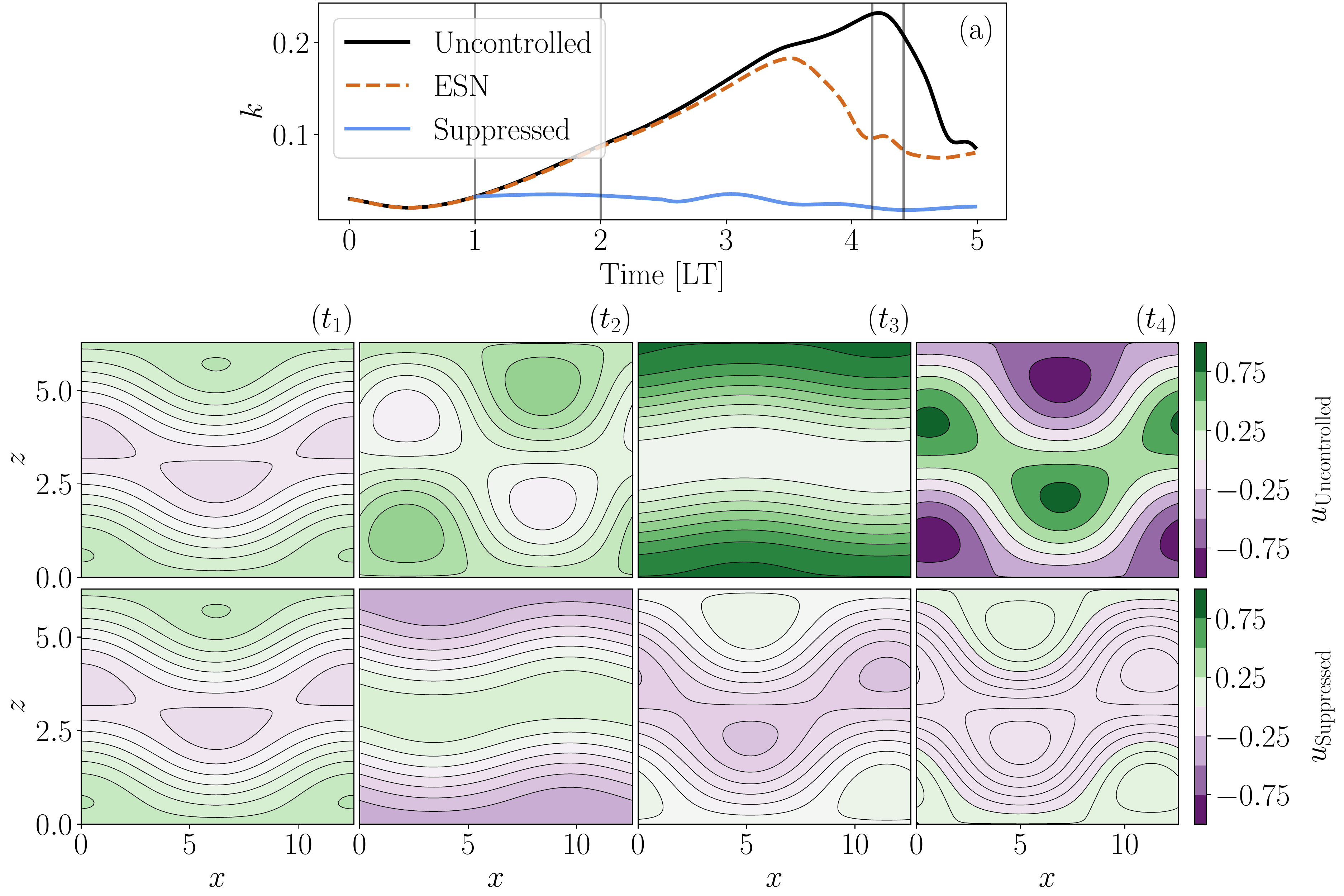}
\caption{Velocity component along $x$ in the midplane with and without control during an extreme event. The times, $(t_i)$, of the flowfields are indicated by the vertical lines in (a). The event is predicted in the interval $[2\mathrm{LTs},3\mathrm{LTs}]$ (PT$=$2LTs), and the control strategy is active from 1LTs to 2.5LTs.}
\label{fig:suppressed}
\end{figure}

In this section, we use the time-accurate prediction of echo state networks (section \ref{sec:prec}) to control and suppress extreme events before they occur. 
We (i) develop a simple and effective control strategy based on the predictions of the networks and (ii) analyse the effects of the control strategy on decreasing the probability of the occurrence of extreme events at different Reynolds numbers.
We deploy the control strategy when the ESN forecasts an extreme event in the future time window lasting 1 Lyapunov time (LT) after the prediction time (PT) (see section \ref{sec:prec}). The control strategy consists of increasing the Reynolds number to 2000 for 1.5 LTs, starting from $\mathrm{PT}-1\mathrm{LT}$ (\cref{fig:control_scheme}).
We choose the value 1.5LTs because it is the average time between the start of the control strategy and the event. 
Physically, increasing the Reynolds corresponds to decreasing the forcing term and the dissipation, which in turn results in a slower change in time of the total energy of the flow \cite{waleffe1997self}. Because we act on the flow before the event happens, reducing the increase of the energy of the flow prevents the kinetic energy from increasing and passing the extreme event threshold.
An example of a controlled extreme event is shown in  \cref{fig:suppressed}. The control strategy is effective and the event is suppressed. This results in significantly lower values of the velocity flowfield in the suppressed case with respect to the uncontrolled case (times $t_3$-$t_4$). Practically, the suppressed time series in panel (a) is obtained by reintegrating the governing equations from 1LT, with Re=2000 from 1 to 2.5LTs and Re=400 from 2.5LTs onwards. The new time series becomes the true data and it is used to monitor the flow to predict future extreme events.
%

\begin{figure}[H]
\centering
\includegraphics[width=.75\textwidth]{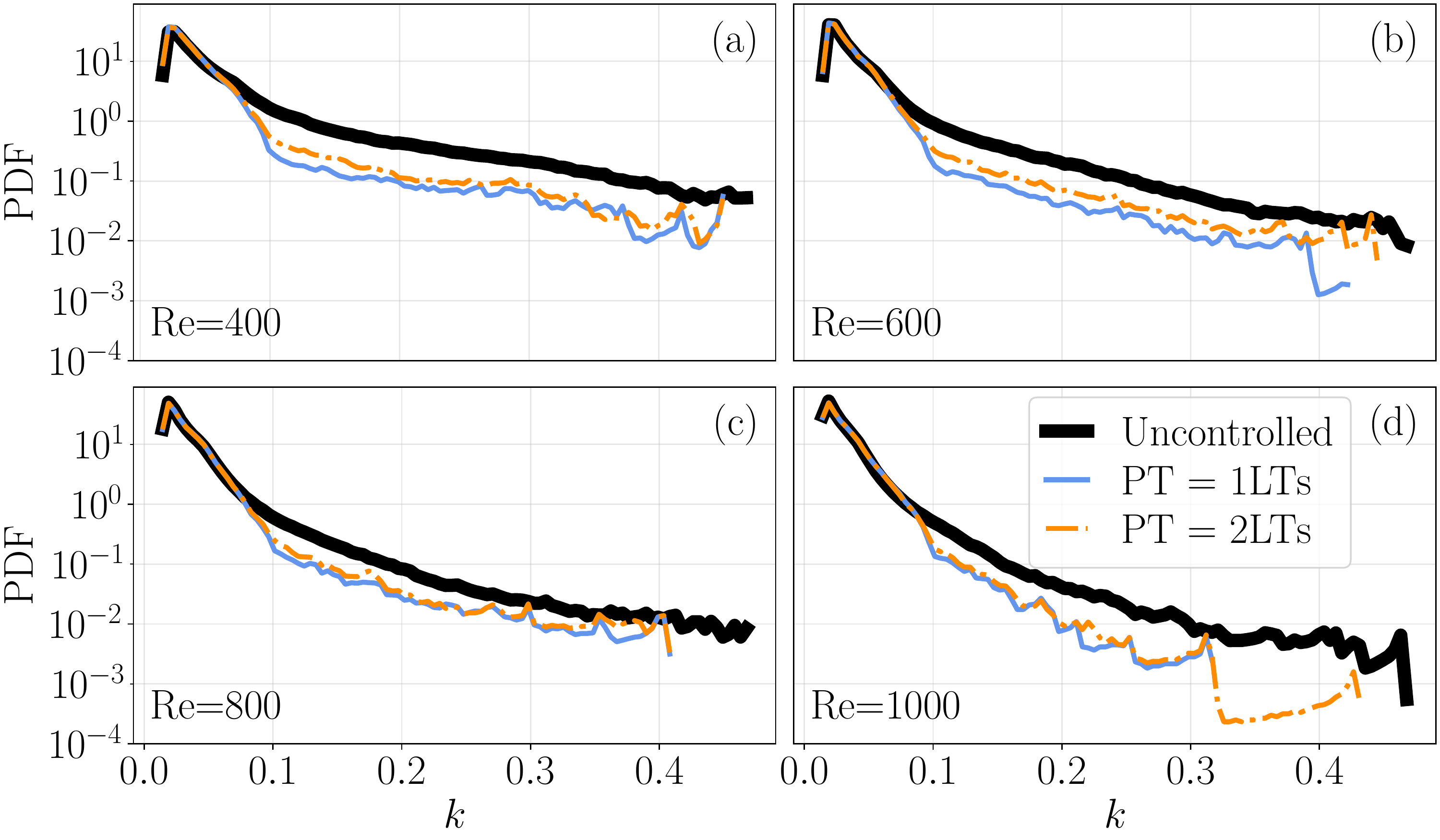}
\caption{Probability Density Function of the kinetic energy with and without control for different prediction times (PT) and Reynolds numbers. }
\label{fig:suppressed1}
\end{figure}

\cref{fig:suppressed1} shows the effect of the control strategy on the average PDF of the kinetic energy for the 2000 neurons network ensemble trained with Recycle Validation at different Reynolds numbers. The tail of the PDF decreases in all cases analysed by up to one order of magnitude for values larger than the extreme event threshold, $k_e=0.1$. This physically means that we are able to predict the vast majority of the events and suppress them. However, the tail does not go to zero for two reasons: (i) not all events are predicted by the networks; (ii) the control strategy is not perfect, i.e., some predicted events are not suppressed. A predicted event is not suppressed if the kinetic energy crosses the extreme event threshold in the interval before the next prediction is made, which is for  $t < 2\mathrm{PT} + 1\mathrm{LT}$. Overall, there is an improvement in the suppression of the tail as the prediction time decreases because the number of false negatives increases with the prediction time (panel (d) in \cref{fig:F}). This is due to the butterfly effect that hinders the time-accurate  prediction for large times. This implies that fewer events are predicted and controlled when the networks try to predict the flow for larger times.

\section{Conclusions}

\label{sec:concl}

We propose a data-driven framework and methodology to predict and control extreme events in nonlinear flows. The data-driven framework is based on echo state networks, which are a type of reservoir computing. 
We analyse a prototypical chaotic shear flow, in which extreme events manifest themselves as sudden bursts of the kinetic energy, which are caused by an intermittent competition between laminar and chaotic dynamics. 
%
%
%

First, we show that a careful hyperparameter optimization is key to the networks' precision, recall, and robustness.
We deploy the Recycle Validation for hyperparameter selection, which is a principled strategy based on a chaotic time scale. The Recycle Validation 
exploits 
Bayesian sampling, it is computationally cheap, and it allows the machine to accurately predicts more extreme events than other hyperparameter-selection techniques do (such as the commonly used Single Shot Validation).
Second, we analyse the networks' capability of time-accurately predicting extreme events before they occur.
We show that the networks correctly predict extreme events up to five Lyapunov times in advance with large precision (up to $98\%$) and recall (up to $99\%)$.
Physically, the networks accurately forecast the incoming attempts of the flow to laminarize and readjust itself through chaotic bursts. 
Third, we analyse the networks' capability of inferring the long-term statistics from relatively-short datasets. 
The networks correctly learn and extrapolate the long-term statistics of the kinetic energy and the velocity from training data with non-converged statistics. 
%
%
Fourth, we leverage on the echo state networks' accurate predictions to control the flow and prevent extreme events from occurring. 
The control strategy is simple and effective. 
The network provides a precursor of the event, which 
activates a control strategy that modifies the Reynolds number to decrease the energy growth rate. 
We reduce the occurrence of extreme events by up to one order of magnitude with respect to the uncontrolled flow, which physically reduces the frequency of the flow laminarization attempts. 
Fifth, by investigating the chaotic shear flow at different Reynolds numbers, we find that the networks are able to predict and control extreme events for a wide range of regimes. This means that the networks performance is robust with respect to changes in the system's physical parameter. 

This paper provides a data-driven framework to predict extreme events in chaotic systems, which can be potentially applied to other nonlinear systems within and beyond fluid mechanics. 
Current and future work is focused on extending the framework to higher-dimensional turbulent flows with a data-compression strategy.

\section*{Code}
The code is available on the github repository \href{https://github.com/MagriLab}{github/MagriLab}.

\section*{Declaration of Interests}
The authors report no conflict of interest. 

\section*{Acknowledgements}
A. Racca is supported by the EPSRC-DTP and the Cambridge
Commonwealth, European \& International Trust under a Cambridge European Scholarship. L. Magri gratefully acknowledges financial support from the ERC Starting Grant PhyCo  949388 and the visiting fellowship at the Technical University of Munich – Institute for Advanced Study, funded by the German Excellence Initiative and the European Union Seventh Framework Programme under grant agreement n. 291763. 
\appendix

\section{Training and validation sets}
\label{sec:train}

In this section, we report additional details regarding the training and validation of echo state networks.
In all the cases analysed, the washout interval consists of 0.5LTs. 
In the Single Shot Validation, the single validation interval consists of 2LTs at the end of the first time series in the ten time series ensemble we use for training (section \ref{sec:MFE}). 
In the Recycle Validation, in the $\mathrm{Re}=400$ case, we use 90 validation intervals, each  of length 2LTs. Subsequent intervals are generated by moving the validation interval forward in time by its own length. Therefore, 90 intervals span the first three time series of the ten time series ensemble we use for training (each time series lasts $\simeq 65 \mathrm{LTs}$, see section \ref{sec:MFE}), so that 30 intervals per time series span $60 \mathrm{LTs}$). In the $\mathrm{Re}=[600,800,1000]$ cases, we use 88 validation intervals, each  of length 2LTs, spanning the first four time series of the ten time series ensemble we use for training. We do so because, due to smaller Lyapunov exponents, the 4000 time units consist of less than $60 \mathrm{LTs}$.


%


\end{document}